\documentclass[12pt,preprint]{aastex}

\slugcomment{The Astronomical Journal, in press}

\shorttitle{HST lightcurves of 20--100 kilometer KBOs}
\shortauthors{Trilling \& Bernstein}

\newcommand\fv{{2000~FV$_{53}$}}
\newcommand\newbf{{2003~BF$_{91}$}}
\newcommand\newbg{{2003~BG$_{91}$}}
\newcommand\newbh{{2003~BH$_{91}$}}

\received{2004 May 4}
\begin{document}

\title{Lightcurves of
20--100 kilometer
Kuiper Belt Objects using the Hubble
Space Telescope\footnote{Based
on observations made with the NASA/ESA Hubble Space Telescope, obtained 
at the Space Telescope Science Institute, which is operated by 
the Association of Universities for Research in Astronomy, Inc., 
under NASA contract NAS 5-26555. 
These observations are associated with program \#9433.}}

\author{David E.\ Trilling\altaffilmark{2}, Gary M.\ Bernstein}
\affil{Department of Physics and Astronomy, University
of Pennsylvania, David Rittenhouse Laboratory,
209 S.\ 33rd St., Philadelphia, PA 19104}
\email{trilling@astro.upenn.edu}
\altaffiltext{2}{Present Address:
Steward Observatory, University of Arizona,
933 N.\ Cherry Avenue, Tucson, AZ 85721;
{\tt trilling@as.arizona.edu}}

\begin{abstract}
We report 
high precision photometry of 
three small and one larger Kuiper
Belt Objects (KBOs) obtained
with the Advanced Camera for Surveys
onboard the Hubble Space Telescope (ACS/HST).
The three small bodies
are the smallest KBOs 
for which lightcurve measurements are available.
\newbf\ has 
a diameter of 20~kilometers (assuming 10\% albedo) and
a 1.09~magnitude, 9.1-hour lightcurve
that is feasibly explained
by the rotation of an elongated,
coherent body that is supported by material strength and best
imagined as an icy outer Solar System
analog to asteroid (243)~Ida.
Two other small KBOs, 
\newbg\ and \newbh\ (diameters 
31~and 18~km, with albedo 10\%), exhibit
an unremarkable lightcurve and 
no detectable photometric variation, respectively.
For the larger KBO \fv\ (116~km diameter,
assuming 10\% albedo) we strongly
detect a
non-sinusoidal periodic (7.5~hours)
brightness variation
with a very small amplitude (0.07~mag).
This KBO
may be nearly spherical,
a result that might not be unusual in the Kuiper
Belt but would be remarkable among outer
Solar System satellites of similar size.


Lightcurves may be caused by variations
in albedo or shape, and
we carry out
a study of possible physical states and bulk densities
under the assumptions of both fluid equilibrium and finite,
non-zero internal friction. 
Under most assumptions, the densities for
the these KBOs are
likely to be in the range 1--2~g~cm$^{-3}$,
and a plausible solution for \fv\ is a
rubble pile of this density that is held slightly out of the
minimum-energy shape by internal friction among constituent blocks
that are relatively small.
Our interpretation of \fv\ as a pulverized but
essentially primordial object and \newbf\
as a collisional fragment is consistent
with models of collisional timescales in the outer
Solar System.
We compile all published KBO lightcurve data to date and 
compare our results to the larger population.


\end{abstract}


\section{Introduction}
\label{introduction}

The Kuiper Belt, a remnant debris disk
that surrounds the planetary realm of our
Solar System,
is a relatively pristine record of the
prevailing conditions during the formation
of the Solar System.
Subsequent evolution has overprinted
this original state such that
present day observations allude to
the combination of
accretion and eons of collisions.
Observations of individual Kuiper
Belt Objects (KBOs) allow
explorations of bodies with
aged, but primordial,
compositions.
Studies of binary KBOs
\citep{veillet02,noll02,osip03,noll04a,noll04b,stans05,stephens05}
provide albedo constraints, usually under a density
assumption,
and may
increase what little is known about
the internal composition and structure of KBOs.

Small Solar System body lightcurves have
been studied for many years, principally
for asteroids \citep{pravec02} and comets
(see, e.g.,
\citet{jewitt91,nalin}).
Lightcurves of small objects are often interpreted as manifestations
of reflections from irregularly shaped objects,
and
lightcurve information has been shown to
correspond well with radar-derived shape
models (see, e.g., \citet{ast3}).
With the continued increase in the number
of known KBOs (and hence bright KBOs) and
access to improving
observing technologies,
KBO lightcurves can now be studied.
Published lightcurve data
exist for 65~KBOs and Centaurs,
with approximately half showing
lightcurves greater than around 0.1~magnitudes
with a typical amplitude around 0.5~magnitudes.
2001~QG$_{298}$ has the largest known
KBO lightcurve amplitude of 1.14~magnitudes \citep{sj04}.
KBO lightcurves are though to imply
either heterogeneous albedo distributions
or else asphericities,
with the extreme case of the latter
potentially being contact binary KBOs \citep{sj04}.
The KBO
lightcurve literature is
tabulated
and analyzed in Section~\ref{densities}.

Here we report 
high precision photometry for four
KBOs observed with the Hubble Space
Telescope (Section~\ref{obs}). The capabilities of the HST and the
extended duration of this study permit, for
the first time, the study of photometric variations of very faint
and
therefore small
KBOs and the detection of very small ($<0.1$~mag)
variations of modest-sized 
KBOs.
Two of these KBOs show clear
periodic variation; a third shows
a somewhat less significant periodic
variation;
and the fourth has no distinguishable
periodic signature
(Section~\ref{analysis}).
We discuss interpretations of this 
data in Sections~\ref{surface}
and~\ref{geophysics} and
the implications in Section~\ref{discussion}.

\section{Observations}
\label{obs}

We have carried out a large (125~orbits)
Hubble Space Telescope/Advanced 
Camera for Surveys (HST/ACS)
observing program to search for very
faint KBOs. The primary results of
this program --- discovery of a substantial deficit
of classical and excited KBOs
at small sizes --- are reported in 
\citet{gmb}. Here we briefly summarize
the relevant technical
details of the 
observations and data reduction
(see \citet{gmb}
for complete discussions).
Our observations were divided into
two epochs, the
``discovery epoch'' (UT 2003 January 26.014--31.341),
in which $55\times400$~s exposures were taken
at each of the six pointings; and the
``recovery epoch'' (2003 February 05.835--09.703),
in which an additional $40\times400$~s exposures
were taken at each pointing.
During the discovery epoch,
a given pointing is sampled sporadically, with
intervals as small as 8~minutes,
over a time span of approximately
24~hours; the pointing is revisited
2~days later with the same sporadic
sampling. Approximately 7~days later,
the entire cycle repeats for the recovery
epoch.
Consequently, we obtained
$\approx95$~independent measurements of each KBO
observed, over a time baseline of
around 12~days, with sampling as fine
as minutes in some cases but with 
windows of several days (or more) in
which no observations of a given KBO
were made.

\citet{gmb} detected three new KBOs
(\newbf , \newbg , and \newbh ) as well as a previously known KBO (\fv
) that was targeted in the observations.  \fv\ was detected with
signal-to-noise ratio $S/N\geq80$ in each of its individual exposures;
hence, the
photometry is quite precise. \newbg, \newbf, and \newbh\ were
discovered with $S/N$ in individual exposures typically 7.5, 2.7, and 2.4,
respectively.  Discovery of the last object required the use of
``digital tracking,'' in which exposures are shifted at rates
corresponding to all valid KBO orbits before summing and searching for
flux peaks that exceed the detection threshold. 

Photometry for each object was extracted by fitting a model of a
moving point source to relevant exposures.
We measured the
point spread function (PSF) for the ACS Wide Field Camera (WFC)
in exposures of a globular cluster field.
In the moving point source model-fitting,
this PSF was smeared before fitting
to the relevant pixels
to account for the
(slight) trailing expected on each
exposure.
We fit the entire stack of
images simultaneously, with the free parameters being the 6
relevant degrees of freedom in the KBO orbit plus an unknown flux for
each exposure.  The best-fit photometry and orbit are thus solved
simultaneously.  

The $S/N$ per exposure for \fv\ is so high that we use a slightly
different approach, allowing the position to be a free parameter on
each exposure rather than forcing positions to obey
a common orbit.  Without
this approach, we find that small (milliarcsecond) errors in the
astrometric solutions 
for the WFC cause excess variance in the flux determinations.
For the fainter three KBOs,
the flux errors due to these $\sim$5~milliarcsecond astrometric
errors are a few
hundredths of a magnitude, well below the noise levels.
The slow brightening due to the decreasing illumination
phase of the KBOs is too small to be detected in our data.

The fitting process produces uncertainties for
each flux measurement.  We
find that the best-fit sinusoidal light curves
give $\chi^2$ per degree of freedom (DOF) near unity for the three faint
bodies (see below),
suggesting that our error estimates are reliable.  The
$\chi^2$ for the best sinusoidal fit
for \fv\ is too high, partly because the
light curve is clearly not sinusoidal (see below), but also because various
systematic effects
(e.g., pointing jitter)
may affect the PSF fitting at the 0.01~mag level.
The formal errors on the magnitudes may also
be underestimated, as is common for very high
$S/N$ photometry.

The midpoint time of each exposure is corrected for light-travel time
from the target.  The time-series photometry for these four
objects is presented in 
Tables~\ref{bgtable} -- \ref{fvtable}.



\begin{deluxetable}{cc}
\tablewidth{0pt}
\tablecaption{Photometry for \newbg}
\tablehead{
\colhead{MJD} &
\colhead{Flux (electrons/sec)}}
\startdata
52666.1767 & 0.3777 $\pm$ 1.8490 \\
52666.1803 & 0.8904 $\pm$ 0.1151 \\
52666.1860 & 0.8166 $\pm$ 0.1134 \\
52666.1918 & 0.7036 $\pm$ 0.1110 \\
52666.1975 & 0.8721 $\pm$ 0.1173 \\
52666.2033 & 0.8127 $\pm$ 0.1168 \\
52666.2421 & 0.6223 $\pm$ 0.0983 \\
52666.2487 & 0.8509 $\pm$ 0.1043 \\
52666.2552 & 0.9279 $\pm$ 0.1058 \\
52666.2618 & 0.8789 $\pm$ 0.1041 \\
52666.2683 & 0.7230 $\pm$ 0.1056 \\
52666.3089 & 0.5674 $\pm$ 0.0968 \\
52666.3154 & 0.8129 $\pm$ 0.1035 \\
52666.3220 & 0.8537 $\pm$ 0.1002 \\
52666.3285 & 0.6708 $\pm$ 0.0992 \\
52666.3351 & 1.0203 $\pm$ 0.1064 \\
52666.5907 & 0.8252 $\pm$ 0.1065 \\
52666.5969 & 0.8402 $\pm$ 0.1053 \\
52666.6031 & 0.8775 $\pm$ 0.1068 \\
52666.6423 & 0.6289 $\pm$ 0.1041 \\
52666.6485 & 0.7175 $\pm$ 0.1025 \\
52666.6549 & 0.5255 $\pm$ 0.0973 \\
52666.6614 & 0.6206 $\pm$ 0.0979 \\
52666.6680 & 0.7295 $\pm$ 0.1008 \\
52666.7101 & 0.6419 $\pm$ 0.0984 \\
52666.7166 & 0.7399 $\pm$ 0.1011 \\
52666.7232 & 0.7426 $\pm$ 0.1021 \\
52666.7297 & 0.8365 $\pm$ 0.1006 \\
52666.7363 & 0.7053 $\pm$ 0.1016 \\
52666.7809 & 0.8550 $\pm$ 0.1031 \\
52666.7874 & 0.8872 $\pm$ 0.1031 \\
52668.8718 & 0.8502 $\pm$ 0.1132 \\
52668.9200 & 0.6952 $\pm$ 0.1042 \\
52668.9263 & 0.4257 $\pm$ 0.1014 \\
52668.9325 & 0.6454 $\pm$ 0.1274 \\
52668.9387 & 0.6640 $\pm$ 0.1064 \\
52668.9872 & 0.8258 $\pm$ 0.1037 \\
52668.9938 & 0.5750 $\pm$ 0.0991 \\
52669.0003 & 0.8541 $\pm$ 0.1020 \\
52669.0444 & 0.6653 $\pm$ 0.0995 \\
52669.0509 & 0.9526 $\pm$ 0.1486 \\
52669.0575 & 0.7976 $\pm$ 0.0989 \\
52669.0640 & 0.8437 $\pm$ 0.1030 \\
52669.0706 & 0.6147 $\pm$ 0.0981 \\
52669.1111 & 0.6653 $\pm$ 0.0975 \\
52669.1176 & 0.8206 $\pm$ 0.1001 \\
52669.2718 & 0.8431 $\pm$ 0.1072 \\
52669.3111 & 0.7061 $\pm$ 0.1018 \\
52669.3173 & 0.6439 $\pm$ 0.1026 \\
52669.3235 & 0.7169 $\pm$ 0.1034 \\
52669.3298 & 0.8750 $\pm$ 0.1046 \\
52669.3362 & 0.8124 $\pm$ 0.1032 \\
52669.3780 & 0.8291 $\pm$ 0.1002 \\
52669.3845 & 0.7899 $\pm$ 0.1001 \\
52669.3911 & 0.9923 $\pm$ 0.1663 \\
52669.3976 & 0.9447 $\pm$ 0.1027 \\
52676.6558 & 0.7321 $\pm$ 0.1039 \\
52676.6620 & 0.9838 $\pm$ 0.1115 \\
52676.6682 & 0.7513 $\pm$ 0.1111 \\
52676.6745 & 0.6800 $\pm$ 0.1075 \\
52676.6807 & 0.7510 $\pm$ 0.1128 \\
52676.7191 & 0.8934 $\pm$ 0.0999 \\
52676.7257 & 0.7270 $\pm$ 0.1011 \\
52676.7322 & 0.9269 $\pm$ 0.1078 \\
52676.7388 & 0.6538 $\pm$ 0.1028 \\
52676.7453 & 0.9019 $\pm$ 0.1111 \\
52676.9227 & 0.6814 $\pm$ 0.1025 \\
52676.9290 & 0.8111 $\pm$ 0.1081 \\
52676.9352 & 0.8000 $\pm$ 0.1074 \\
52676.9414 & 0.8414 $\pm$ 0.1101 \\
52676.9477 & 0.8661 $\pm$ 0.1185 \\
52676.9861 & 0.9424 $\pm$ 0.0991 \\
52676.9926 & 0.7074 $\pm$ 0.0995 \\
52676.9992 & 0.8667 $\pm$ 0.1087 \\
52677.0057 & 0.7761 $\pm$ 0.1061 \\
52677.0123 & 0.8983 $\pm$ 0.1080 \\
52678.3383 & 0.7229 $\pm$ 0.1089 \\
52678.3445 & 0.9430 $\pm$ 0.1145 \\
52678.3876 & 0.7263 $\pm$ 0.1002 \\
52678.3938 & 1.0875 $\pm$ 0.1196 \\
52678.4000 & 0.6840 $\pm$ 0.1206 \\
52678.4064 & 0.8274 $\pm$ 0.1095 \\
52678.4130 & 0.7787 $\pm$ 0.1031 \\
52678.4545 & 0.6907 $\pm$ 0.0936 \\
52678.4610 & 0.8603 $\pm$ 0.1042 \\
52678.4676 & 0.6253 $\pm$ 0.1037 \\
52678.6580 & 0.7266 $\pm$ 0.1015 \\
52678.6643 & 0.7281 $\pm$ 0.1033 \\
52678.6705 & 0.7538 $\pm$ 0.1104 \\
52678.6767 & 0.7256 $\pm$ 0.1100 \\
52678.6829 & 0.7295 $\pm$ 0.1115 \\
52678.7215 & 1.0322 $\pm$ 0.1016 \\
52678.7280 & 0.8498 $\pm$ 0.1042 \\
52678.7346 & 0.8373 $\pm$ 0.1087 \\
52678.7411 & 0.7195 $\pm$ 0.1072 \\
52678.7477 & 0.8476 $\pm$ 0.1090 \\
\enddata
\tablecomments{Fluxes and MJD are reported
for individual exposures.
The zero points (the magnitude of a star
that produces a count rate of 1~electron per second in the given filter)
for ACS/WFC 
F606W are 26.486~in
the AB magnitude system
and 26.655~in the ST~magnitude system.}
\label{bgtable}
\end{deluxetable}

\begin{deluxetable}{cc}
\tablewidth{0pt}
\tablecaption{Photometry for \newbf }
\tablehead{
\colhead{MJD} &
\colhead{Flux (electrons/sec)}}
\startdata
52665.8549 & -1.2230 $\pm$ 1.8355 \\
52665.8585 & 0.2728 $\pm$ 0.1085 \\
52665.9086 & 0.3598 $\pm$ 0.1063 \\
52665.9143 & 0.3964 $\pm$ 0.1044 \\
52665.9201 & 0.4106 $\pm$ 0.1024 \\
52665.9764 & 0.3061 $\pm$ 0.1063 \\
52665.9826 & 0.3144 $\pm$ 0.0944 \\
52665.9891 & 0.3846 $\pm$ 0.1639 \\
52666.0438 & 0.3133 $\pm$ 0.0930 \\
52666.0503 & 0.2163 $\pm$ 0.0939 \\
52666.0569 & 0.2491 $\pm$ 0.0936 \\
52666.0979 & -0.1329 $\pm$ 0.0888 \\
52666.1044 & 0.1358 $\pm$ 0.0934 \\
52666.1110 & 0.1404 $\pm$ 0.0930 \\
52666.1175 & 0.1398 $\pm$ 0.0919 \\
52666.1241 & 0.1718 $\pm$ 0.0919 \\
52668.6357 & 0.1921 $\pm$ 0.0927 \\
52668.6419 & 0.4871 $\pm$ 0.0979 \\
52668.6482 & 0.3701 $\pm$ 0.0969 \\
52668.6544 & 0.3791 $\pm$ 0.0958 \\
52668.6606 & 0.4132 $\pm$ 0.1010 \\
52668.7010 & 0.3538 $\pm$ 0.0906 \\
52668.7076 & 0.3290 $\pm$ 0.0909 \\
52668.7141 & 0.4755 $\pm$ 0.0947 \\
52668.7207 & 0.3715 $\pm$ 0.0925 \\
52668.7272 & 0.2321 $\pm$ 0.0935 \\
52668.7715 & 0.2818 $\pm$ 0.0922 \\
52668.7781 & 0.0750 $\pm$ 0.0892 \\
52668.7846 & 0.3555 $\pm$ 0.0922 \\
52668.7912 & 0.0323 $\pm$ 0.0905 \\
52668.8411 & 0.1561 $\pm$ 0.0894 \\
52669.1217 & 0.1626 $\pm$ 0.0946 \\
52669.1279 & -0.0425 $\pm$ 0.0939 \\
52669.1668 & 0.1551 $\pm$ 0.0924 \\
52669.1730 & 0.2126 $\pm$ 0.0944 \\
52669.1792 & 0.1463 $\pm$ 0.0935 \\
52669.1856 & 0.2512 $\pm$ 0.0919 \\
52669.1922 & 0.1659 $\pm$ 0.0912 \\
52669.2337 & 0.0893 $\pm$ 0.0883 \\
52669.2402 & 0.4273 $\pm$ 0.0926 \\
52669.2468 & 0.2239 $\pm$ 0.0915 \\
52676.5115 & 0.4058 $\pm$ 0.0969 \\
52676.5177 & 0.3830 $\pm$ 0.0976 \\
52676.5240 & 0.3819 $\pm$ 0.1085 \\
52676.5302 & 0.3266 $\pm$ 0.1008 \\
52676.5364 & 0.4167 $\pm$ 0.1038 \\
52676.5748 & 0.4739 $\pm$ 0.0897 \\
52676.5814 & 0.4051 $\pm$ 0.0933 \\
52676.5879 & 0.4159 $\pm$ 0.0993 \\
52676.5945 & 0.4730 $\pm$ 0.1004 \\
52676.6010 & 0.5572 $\pm$ 0.1017 \\
52676.7785 & 0.1983 $\pm$ 0.0918 \\
52676.7847 & 0.0316 $\pm$ 0.0960 \\
52676.7909 & 0.1043 $\pm$ 0.0975 \\
52676.7972 & 0.0684 $\pm$ 0.0969 \\
52676.8034 & 0.1977 $\pm$ 0.1037 \\
52676.8418 & 0.4073 $\pm$ 0.0889 \\
52676.8484 & 0.3386 $\pm$ 0.1067 \\
52676.8549 & 0.3457 $\pm$ 0.0966 \\
52676.8615 & 0.5215 $\pm$ 0.1140 \\
52676.8680 & 0.2471 $\pm$ 0.0982 \\
52678.1798 & 0.3723 $\pm$ 0.0975 \\
52678.1860 & 0.1812 $\pm$ 0.0978 \\
52678.1923 & 0.4134 $\pm$ 0.1014 \\
52678.1985 & 0.1108 $\pm$ 0.0984 \\
52678.2047 & 0.2741 $\pm$ 0.1021 \\
52678.2435 & 0.1248 $\pm$ 0.0841 \\
52678.2500 & 0.1667 $\pm$ 0.0931 \\
52678.2566 & 0.1432 $\pm$ 0.0968 \\
52678.2631 & 0.1704 $\pm$ 0.0949 \\
52678.3102 & 0.2398 $\pm$ 0.0885 \\
52678.4718 & 0.3098 $\pm$ 0.1039 \\
52678.5103 & 0.2805 $\pm$ 0.0927 \\
52678.5165 & 0.3448 $\pm$ 0.0998 \\
52678.5227 & 0.4797 $\pm$ 0.1029 \\
52678.5290 & 0.2635 $\pm$ 0.1016 \\
52678.5353 & 0.1256 $\pm$ 0.1947 \\
52678.5772 & 0.1641 $\pm$ 0.0879 \\
52678.5837 & 0.1143 $\pm$ 0.0920 \\
52678.5903 & 0.2215 $\pm$ 0.0970 \\
52678.5968 & -0.0663 $\pm$ 0.0972 \\
\enddata
\tablecomments{Fluxes and MJD are reported
for individual exposures. Zero points are given
in Table~\ref{bgtable}.}
\label{bftable}
\end{deluxetable}

\begin{deluxetable}{cc}
\tablewidth{0pt}
\tablecaption{Photometry for \newbh }
\tablehead{
\colhead{MJD} &
\colhead{Flux (electrons/sec)}}
\startdata
52666.1635 & 0.7434 $\pm$ 1.8236 \\
52666.1671 & 0.1931 $\pm$ 0.1010 \\
52666.1728 & 0.1785 $\pm$ 0.0956 \\
52666.1786 & 0.2867 $\pm$ 0.1315 \\
52666.1843 & 0.1527 $\pm$ 0.0918 \\
52666.1901 & 0.3753 $\pm$ 0.1033 \\
52666.2289 & 0.1168 $\pm$ 0.0864 \\
52666.2355 & 0.1277 $\pm$ 0.0861 \\
52666.2420 & 0.2216 $\pm$ 0.0892 \\
52666.2486 & 0.1254 $\pm$ 0.0832 \\
52666.2551 & 0.0832 $\pm$ 0.0865 \\
52666.2956 & 0.3803 $\pm$ 0.0973 \\
52666.3022 & 0.2458 $\pm$ 0.0891 \\
52666.3087 & 0.2723 $\pm$ 0.0894 \\
52666.3153 & 0.2331 $\pm$ 0.0863 \\
52666.3218 & 0.2510 $\pm$ 0.0895 \\
52666.5775 & -0.0319 $\pm$ 0.0869 \\
52666.5837 & 0.1558 $\pm$ 0.0898 \\
52666.5899 & 0.2509 $\pm$ 0.0931 \\
52666.6291 & 0.3299 $\pm$ 0.0937 \\
52666.6353 & 0.2186 $\pm$ 0.0926 \\
52666.6417 & 0.2202 $\pm$ 0.0894 \\
52666.6482 & 0.2146 $\pm$ 0.0862 \\
52666.6548 & 0.2562 $\pm$ 0.2347 \\
52666.6968 & 0.2091 $\pm$ 0.0880 \\
52666.7034 & 0.2180 $\pm$ 0.0893 \\
52666.7099 & 0.1572 $\pm$ 0.0883 \\
52666.7165 & 0.0368 $\pm$ 0.0821 \\
52666.7231 & -0.0395 $\pm$ 0.0849 \\
52666.7677 & 0.2357 $\pm$ 0.0869 \\
52666.7742 & 0.2447 $\pm$ 0.0879 \\
52668.8586 & 0.2600 $\pm$ 0.0985 \\
52668.9068 & 0.2271 $\pm$ 0.0900 \\
52668.9130 & 0.2700 $\pm$ 0.0923 \\
52668.9193 & 0.0959 $\pm$ 0.0956 \\
52668.9255 & 0.1183 $\pm$ 0.0926 \\
52668.9740 & 0.1358 $\pm$ 0.0857 \\
52668.9806 & 0.1967 $\pm$ 0.0888 \\
52668.9871 & 0.2410 $\pm$ 0.0985 \\
52669.0311 & 0.2483 $\pm$ 0.0881 \\
52669.0377 & 0.2852 $\pm$ 0.0869 \\
52669.0442 & 0.2026 $\pm$ 0.0867 \\
52669.0508 & 0.2812 $\pm$ 0.0878 \\
52669.0573 & 0.2010 $\pm$ 0.0903 \\
52669.0979 & 0.2965 $\pm$ 0.0861 \\
52669.1044 & -0.0006 $\pm$ 0.0820 \\
52669.2586 & 0.1599 $\pm$ 0.0907 \\
52669.2979 & 0.2009 $\pm$ 0.0997 \\
52669.3041 & 0.3301 $\pm$ 0.1140 \\
52669.3103 & 0.3237 $\pm$ 0.0934 \\
52669.3166 & 0.0025 $\pm$ 0.0900 \\
52669.3229 & 0.3061 $\pm$ 0.0898 \\
52669.3648 & 0.0673 $\pm$ 0.1574 \\
52669.3713 & 0.4892 $\pm$ 0.0910 \\
52669.3779 & 0.1185 $\pm$ 0.0955 \\
52669.3844 & 0.2819 $\pm$ 0.0871 \\
52676.6426 & 0.0388 $\pm$ 0.0911 \\
52676.6488 & 0.0465 $\pm$ 0.0947 \\
52676.6550 & 0.4138 $\pm$ 0.1055 \\
52676.6613 & 0.2930 $\pm$ 0.1005 \\
52676.6675 & 0.2630 $\pm$ 0.1042 \\
52676.7059 & 0.2449 $\pm$ 0.0892 \\
52676.7125 & 0.3346 $\pm$ 0.0945 \\
52676.7190 & 0.2906 $\pm$ 0.0972 \\
52676.7256 & 0.3516 $\pm$ 0.0969 \\
52676.7321 & 0.2813 $\pm$ 0.0990 \\
52676.9095 & 0.1801 $\pm$ 0.0925 \\
52676.9158 & 0.0725 $\pm$ 0.0950 \\
52676.9220 & 0.2895 $\pm$ 0.1006 \\
52676.9282 & 0.2319 $\pm$ 0.0981 \\
52676.9345 & 0.3169 $\pm$ 0.1102 \\
52676.9729 & 0.0706 $\pm$ 0.0834 \\
52676.9794 & 0.2492 $\pm$ 0.0944 \\
52676.9860 & 0.1838 $\pm$ 0.0960 \\
52676.9925 & 0.2761 $\pm$ 0.0977 \\
52676.9991 & 0.1273 $\pm$ 0.0949 \\
52678.3251 & 0.2235 $\pm$ 0.1003 \\
52678.3313 & 0.4014 $\pm$ 0.1055 \\
52678.3744 & 0.2857 $\pm$ 0.0909 \\
52678.3806 & 0.0887 $\pm$ 0.0906 \\
52678.3868 & 0.1898 $\pm$ 0.0991 \\
52678.3932 & 0.3169 $\pm$ 0.0988 \\
52678.3998 & 0.1556 $\pm$ 0.0962 \\
52678.4413 & 0.0981 $\pm$ 0.0858 \\
52678.4478 & 0.2142 $\pm$ 0.0912 \\
52678.4544 & 0.2360 $\pm$ 0.1000 \\
52678.6448 & 0.1798 $\pm$ 0.0921 \\
52678.6511 & 0.2354 $\pm$ 0.0960 \\
52678.6573 & 0.2468 $\pm$ 0.0982 \\
52678.6635 & 0.2584 $\pm$ 0.1009 \\
52678.6697 & 0.0378 $\pm$ 0.1025 \\
52678.7083 & 0.2183 $\pm$ 0.0870 \\
52678.7148 & 0.1856 $\pm$ 0.0939 \\
52678.7214 & 0.4244 $\pm$ 0.1015 \\
52678.7279 & 0.2463 $\pm$ 0.0964 \\
52678.7345 & 0.2580 $\pm$ 0.0976 \\
\enddata
\tablecomments{Fluxes and MJD are reported
for individual exposures.
Zero points are given
in Table~\ref{bgtable}.}
\label{bhtable}
\end{deluxetable}

\begin{deluxetable}{cc}
\tablewidth{0pt}
\tablecaption{Photometry for \fv }
\tablehead{
\colhead{MJD} &
\colhead{Flux (electrons/sec)}}
\startdata
52664.8329 & 21.1105 $\pm$ 0.3022 \\
52664.8386 & 21.9243 $\pm$ 0.3048 \\
52664.8444 & 21.3123 $\pm$ 0.3072 \\
52664.8917 & 20.4094 $\pm$ 0.2961 \\
52664.9005 & 19.8693 $\pm$ 0.2640 \\
52664.9093 & 19.5699 $\pm$ 0.2652 \\
52664.9619 & 19.8776 $\pm$ 0.2658 \\
52664.9760 & 19.7707 $\pm$ 0.2660 \\
52665.0304 & 20.9440 $\pm$ 0.2697 \\
52665.0370 & 20.7644 $\pm$ 0.2685 \\
52665.0435 & 19.7313 $\pm$ 0.2649 \\
52665.0975 & 20.6862 $\pm$ 0.2681 \\
52665.1040 & 20.5612 $\pm$ 0.2681 \\
52665.4864 & 20.5181 $\pm$ 0.2818 \\
52665.4926 & 21.4516 $\pm$ 0.2869 \\
52665.4988 & 21.7286 $\pm$ 0.2896 \\
52665.5051 & 21.1291 $\pm$ 0.2799 \\
52665.5113 & 21.6956 $\pm$ 0.2869 \\
52665.5514 & 19.3965 $\pm$ 0.2615 \\
52665.5580 & 19.8271 $\pm$ 0.2642 \\
52665.5645 & 20.3042 $\pm$ 0.2657 \\
52665.5711 & 19.8380 $\pm$ 0.2648 \\
52665.5776 & 20.3634 $\pm$ 0.2680 \\
52665.6174 & 19.5880 $\pm$ 0.2628 \\
52665.6240 & 19.7145 $\pm$ 0.2628 \\
52665.6305 & 19.6671 $\pm$ 0.2622 \\
52665.6371 & 19.7291 $\pm$ 0.2649 \\
52667.9073 & 20.3999 $\pm$ 0.2793 \\
52667.9135 & 20.2463 $\pm$ 0.2793 \\
52667.9687 & 20.9419 $\pm$ 0.2810 \\
52667.9751 & 20.5474 $\pm$ 0.2664 \\
52668.0297 & 20.7742 $\pm$ 0.2681 \\
52668.0362 & 20.8846 $\pm$ 0.2734 \\
52668.0428 & 21.6254 $\pm$ 0.2737 \\
52668.0859 & 19.3086 $\pm$ 0.2580 \\
52668.0925 & 19.4897 $\pm$ 0.2580 \\
52668.0990 & 19.6907 $\pm$ 0.2616 \\
52668.1056 & 19.4131 $\pm$ 0.2591 \\
52668.1121 & 20.2521 $\pm$ 0.2683 \\
52668.1527 & 20.4820 $\pm$ 0.2654 \\
52668.4223 & 19.5664 $\pm$ 0.2696 \\
52668.4285 & 19.9758 $\pm$ 0.2795 \\
52668.4347 & 19.4178 $\pm$ 0.2700 \\
52668.4409 & 20.3281 $\pm$ 0.2761 \\
52668.4472 & 19.9986 $\pm$ 0.2764 \\
52668.4863 & 20.8141 $\pm$ 0.2684 \\
52668.4928 & 20.8260 $\pm$ 0.2695 \\
52668.4994 & 21.1886 $\pm$ 0.2688 \\
52668.5059 & 21.1099 $\pm$ 0.2712 \\
52668.5125 & 21.0477 $\pm$ 0.2707 \\
52675.6487 & 19.0706 $\pm$ 0.2783 \\
52675.6550 & 18.6601 $\pm$ 0.2747 \\
52675.6934 & 20.4433 $\pm$ 0.2740 \\
52675.6996 & 20.6350 $\pm$ 0.2771 \\
52675.7058 & 20.1656 $\pm$ 0.2797 \\
52675.7122 & 19.3632 $\pm$ 0.2674 \\
52675.7188 & 19.0083 $\pm$ 0.2662 \\
52675.7669 & 20.6004 $\pm$ 0.2704 \\
52675.7734 & 20.1532 $\pm$ 0.2717 \\
52676.4308 & 20.2407 $\pm$ 0.2741 \\
52676.4370 & 20.4120 $\pm$ 0.2780 \\
52676.4495 & 20.1571 $\pm$ 0.2835 \\
52676.4557 & 20.3265 $\pm$ 0.2856 \\
52676.4945 & 20.7407 $\pm$ 0.2671 \\
52676.5011 & 19.9136 $\pm$ 0.2653 \\
52676.5076 & 20.0358 $\pm$ 0.2729 \\
52676.5142 & 19.8045 $\pm$ 0.2720 \\
52676.5207 & 19.8140 $\pm$ 0.2751 \\
52677.6993 & 20.6344 $\pm$ 0.2799 \\
52677.7055 & 19.7917 $\pm$ 0.2806 \\
52677.7118 & 20.3615 $\pm$ 0.2884 \\
52677.7180 & 19.4943 $\pm$ 0.2786 \\
52677.7242 & 20.0349 $\pm$ 0.2823 \\
52677.7627 & 20.1795 $\pm$ 0.2608 \\
52677.7692 & 20.4331 $\pm$ 0.2695 \\
52677.7758 & 20.4282 $\pm$ 0.2926 \\
52677.7823 & 20.1118 $\pm$ 0.2813 \\
52677.7889 & 19.8429 $\pm$ 0.2702 \\
52677.9725 & 20.9603 $\pm$ 0.2831 \\
52677.9787 & 22.3236 $\pm$ 0.2950 \\
52677.9849 & 21.5780 $\pm$ 0.2932 \\
52677.9912 & 20.7065 $\pm$ 0.2891 \\
52678.0296 & 20.6365 $\pm$ 0.2662 \\
52678.0362 & 20.1547 $\pm$ 0.2707 \\
52678.0427 & 20.2353 $\pm$ 0.2789 \\
52678.0493 & 19.5048 $\pm$ 0.2773 \\
52678.0558 & 20.1629 $\pm$ 0.2722 \\
\enddata
\tablecomments{Fluxes and MJD are reported
for individual exposures, excluding
measurements from 
from 9~exposures that exhibit
cosmic rays or image defects close to
the KBO image, 
Zero points are given
in Table~\ref{bgtable}.}
\label{fvtable}
\end{deluxetable}

\section{Analysis}
\label{analysis}

\begin{figure}
\plotone{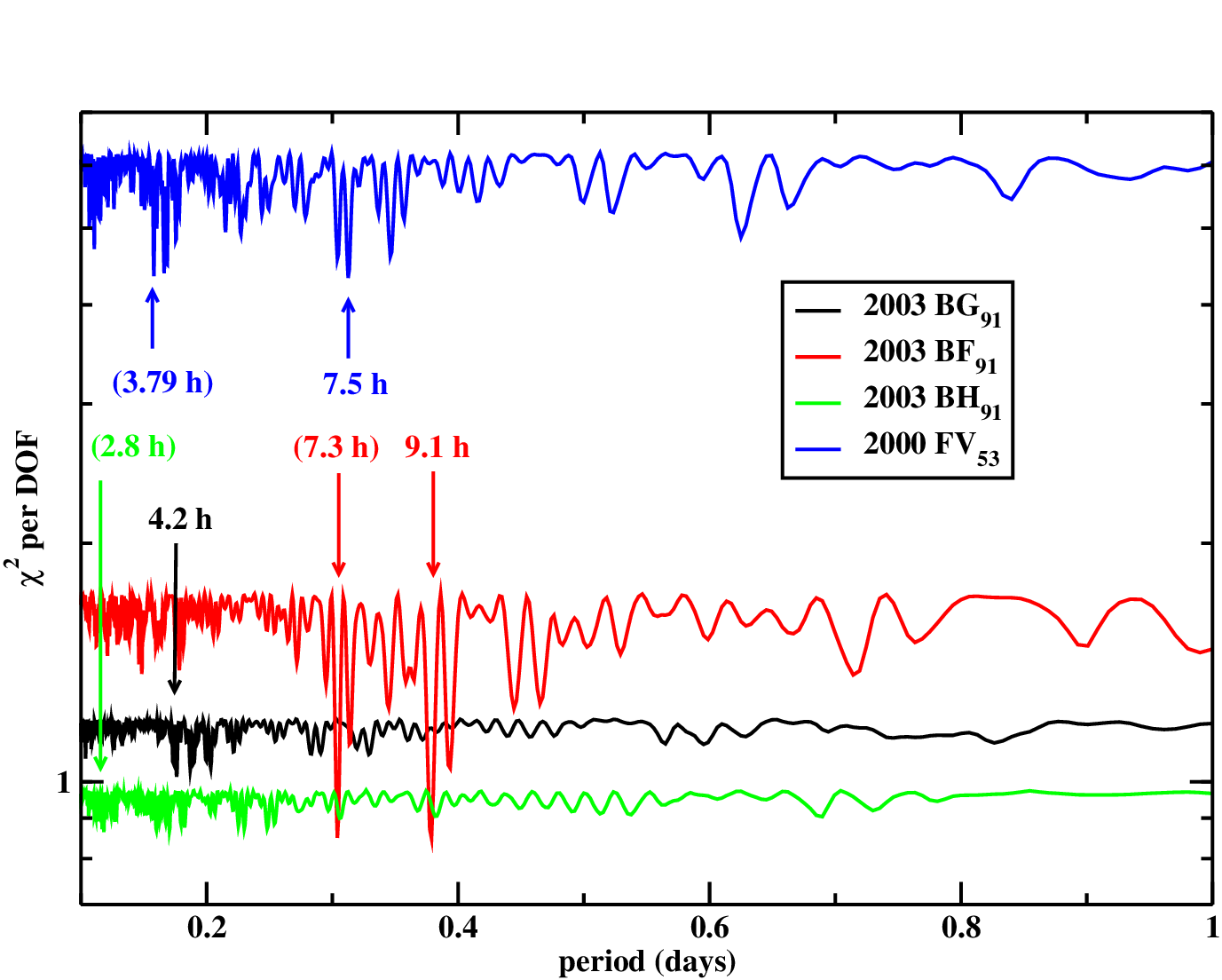}
\caption[]{Periodogram for observed
data. The best-fit periods (given in Table~\ref{solutions}
and shown in Figure~\ref{phased}) are marked
with the arrows, as are other peaks of 
note, in parentheses.
\newbg\ (black) has a suite of solutions,
of which the best has a period of 
4.2~hours; the others correspond to
periods of 4.5~and 4.9~hours.
\newbf\ (red)
has a secondary, non-resonant peak
at 7.3~hours that is nearly as
satisfactory a fit as the best fit.
The best-fit solution for
\newbh\ (green) is not significant and
is therefore marked in parentheses.
The best-fit period for
\fv\ (blue) is 7.5~hours;
a solution that is nearly as
significant
is found with
a period of 3.79~hours, almost
exactly half the best-fit solution (see text).
}

\label{period}
\end{figure}

\begin{figure}
\plotone{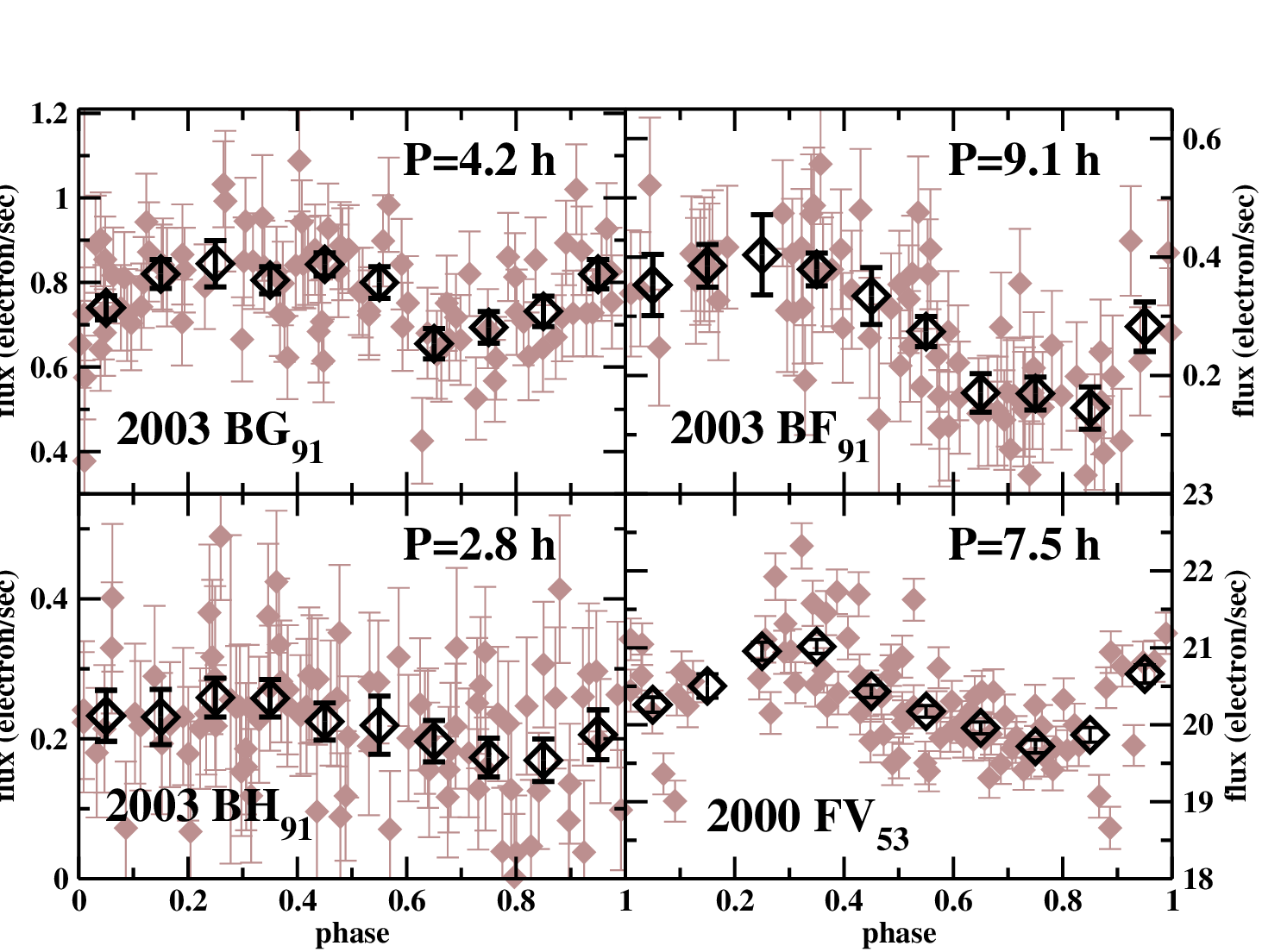}
\caption[]{Data folded at best-fit periods
(as labeled) as a function of 
phase. The pale, filled diamonds show
the individual photometric measurements;
the large, open symbols show the 
observed data binned
with bins 0.1~phase units wide.
The lightcurve amplitudes
for \newbg, \newbf, and
\fv\ are listed in
Table~\ref{solutions}.
The best-fit solution for
\newbh, with signifance
of only 46\%, is
a period of 2.8~hours and
amplitude of 0.42~magnitudes.}
\label{phased}
\end{figure}

We analyze the time-series photometry
of these four KBOs for periodic variations
by searching
for the best-fit sinusoid
lightcurve variation to the observed data.
We search periods $P\ge0.1$~days with uniform steps in frequency
($1/P$) of 0.01~days$^{-1}$;
shorter periods would tend to be badly aliased
by the 96-minute HST orbital period.
For each of the newly discovered KBOs, we fit
all the individual photometric measurements.
Because the $S/N$ for each
individual \fv\ measurement is so high,
we exclude
flux values from 9~exposures that exhibit
cosmic rays or image defects close to
the KBO image, leaving 87~valid flux
measurements.
The resultant periodograms are
plotted in Figure~\ref{period}.
The best-fit solutions are
plotted in Figure~\ref{phased} as a function
of phase, and listed in
Table~\ref{solutions} (together with
estimated uncertainties in derived amplitude).
None of the three new KBOs shows any evidence for 
double-peaked lightcurves, but
the sinusoidal periods we derive could easily
represent a half-rotation (as
in the aspherical case ---
Section~\ref{shapes}) rather than a complete rotation
period (as in the
albedo case --- Section~\ref{albedo}).
The best-fit period solution for \fv\ appears double-peaked,
though half this period may also be a valid
solution (see below).

\begin{deluxetable}{lcccccc}
\tablewidth{0pt}
\tablecaption{Best-fit solutions for KBO lightcurves}
\tablehead{
\colhead{Object} & 
\colhead{mean} & 
\colhead{diameter\tablenotemark{a}} & 
\colhead{Period} & 
\colhead{Amplitude} & 
\colhead{Amplitude} & 
\colhead{Significance\tablenotemark{b}} \\
\colhead{} &
\colhead{F606W} &
\colhead{} &
\colhead{} &
\colhead{} &
\colhead{uncertainty} & 
\colhead{} \\
\colhead{} &
\colhead{mag} &
\colhead{} &
\colhead{} &
\colhead{} & 
\colhead{} &
\colhead{} \\
\colhead{} &
\colhead{({\tt STMAG})} &
\colhead{(km)} &
\colhead{(hrs)} &
\colhead{(mag)} &
\colhead{(mag)} &
\colhead{(\%)}
}
\startdata
\newbg & 26.95$\pm$0.02 & 31 (48) & 4.2 & 0.18 & 0.075 & 90 \\
\newbf & 28.15$\pm$0.04 & 20 (31) & 9.1 &1.09&0.25 & $>$99 \\
\newbh\tablenotemark{c} & 28.38$\pm$0.05 & 18 (28) & \nodata & $<$0.15 & \nodata &  \nodata \\
\fv    & 23.41$\pm$0.01 & 116 (183) & 7.5 & 0.07 & 0.02 &  $>$99 \\
\enddata
\tablenotetext{a}{From \citet{gmb} and erratum.
Assumes 10\% (and, in parentheses, 4\%)
albedos and spherical bodies.}
\tablenotetext{b}{Significance of the best-fit solution
compared to 100~randomizations of the data; that is, the
number of the 100~random data trials that provide worse $\chi^2$/DOF
than the observed data (see Section~\ref{analysis}
and Figure~\ref{random}).}
\tablenotetext{c}{Here
we list
our derived upper limit that corresponds to
no detection of periodic variation (Section~\ref{newbh}).
}
\label{solutions}
\end{deluxetable}

\begin{figure}
\plotone{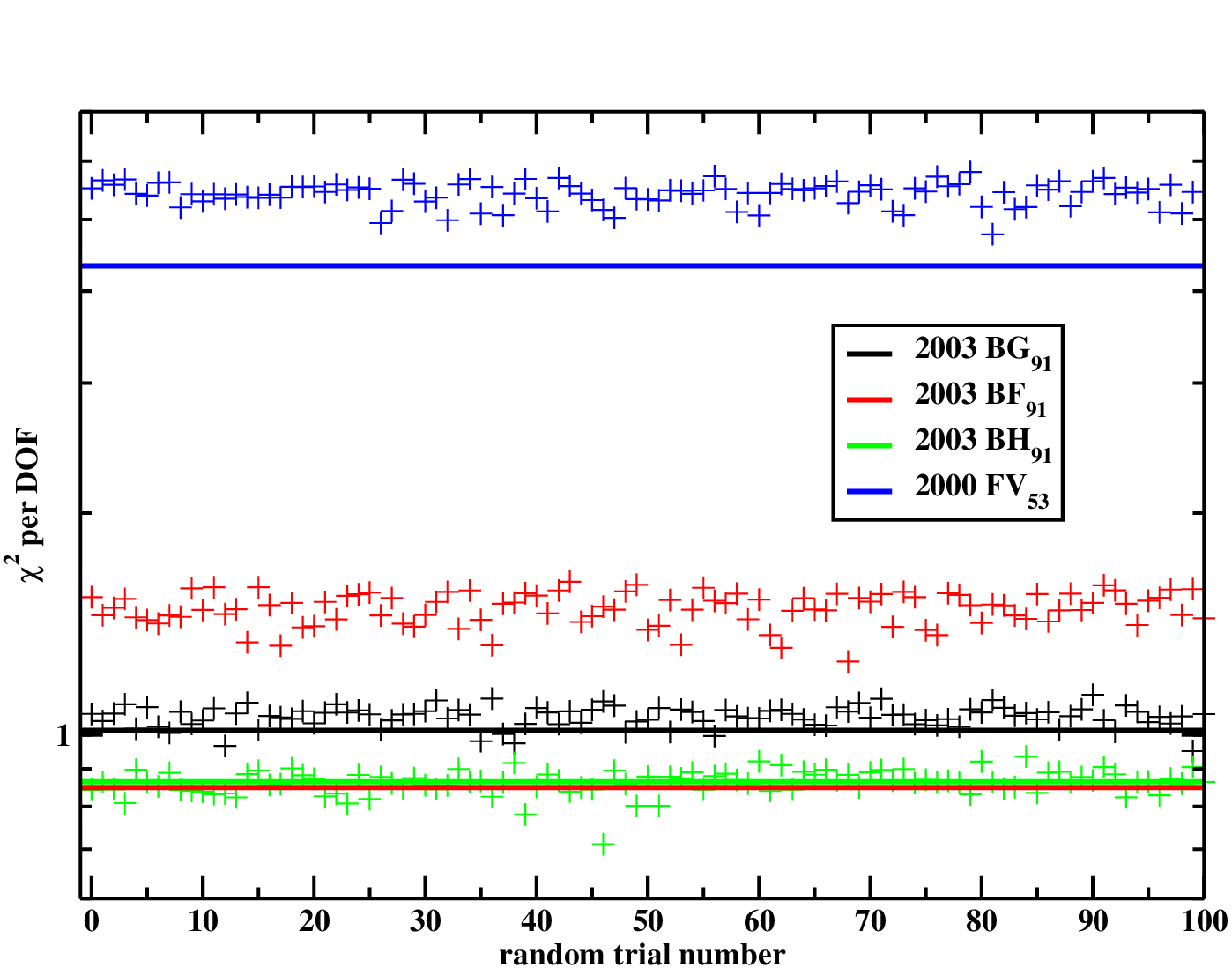}
\caption[]{
Significances of the
best-fit solutions to (true) observed data (solid lines).
We also show
the $\chi^2$ per degree of freedom (DOF) for the best
sinusoidal fit to each of
100 randomizations of each KBO's true observed data (crosses).
For \newbf\ and \fv ,
the true observed data are clearly significantly
better
fit by a sinusoid than are
randomized versions of those data.
The best fit for the true observed
\newbg\ data is marginally significant,
and the \newbh\ true observed
data is not better
fit than randomized data.
Note that the vertical axis is 
logarithmic.
}
\label{random}
\end{figure}

To determine the significance of each best-fit solution,
we randomly scramble the time tags of the flux
measurements for a given object and repeat the
search for a best-fit sinusoid.
For each KBO we fit 100~randomizations of the
data, with the resulting best $\chi^2$/DOF
of each trial plotted in
Figure~\ref{random}. 
Best-fit solutions to randomized \newbf\ and \fv\ data are clearly
less good than the best-fit solution to observed data at $>$99\%
confidence in both cases.  The best-fit solution to the observed
\newbg\ data is marginally significant (90\% confidence level),
while the
best fit for \newbh\ data is no better than the best fits to
randomized data.

%

\subsection{\newbg }
\label{newbg}

\newbg\ has a best-fit sinusoidal
solution with a period of 4.2~hours
and amplitude of~0.18 magnitudes.
Similar, but less good, solutions
are found for 4.5, 4.6, and 4.9~hours;
these periods are not obviously aliases
of each other.
In the discussion that follows, we
refer only to the best-fit period of
4.2~hours and, regardless
of period solution,
draw no significant conclusions
about the internal properties 
of KBOs from this body's lightcurve.

\subsection{\newbf }
\label{newbf}

The periodogram for \newbf\ shows
two clear solutions that are nearly equivalently
good fits: 9.1~hours and, secondarily, 7.3~hours (Figure~\ref{period}).
Both solutions have amplitudes of 1.09~magnitudes.
The secondary peak is non-resonant
with the best fit, i.e., not an
obvious harmonic of the best-fit period,
and is also quite significant compared to
the randomized data.
We therefore searched further for a best-fit solution
that consisted of 
two independent sinusoids
with independent phases, amplitudes,
and periods, though the periods were restricted
to a small range around each of
best fits
derived above.
Formally, the $\chi^2$ improves
significantly through allowing a 
two-sine fit, but the data may not
warrant attaching too much importance
to a multiple rotation pole interpretation.
In the following analysis, we use
the single-sinusoid better-fitting 9.1~hour
period. None of our conclusions depend upon
the choice between the two best-fit
single-sinusoid periods, nor particularly
on the choice of single- over double-sinusoid
fit.

\subsection{\newbh }
\label{newbh}

The best-fit sinusoid solution to the 
photometry of \newbh\ has a period
of 2.8~hours and amplitude of 
0.42~magnitudes. However, the significance
of this solution is only 46\% when
compared to 100~randomized trials of 
the \newbh\ data. We therefore conclude
that we
failed to detect significant periodic
variability for \newbh.
To place an upper limit on the amplitude
of an undetected periodic variation, we
augmented the \newbh\ data with synthetic
lightcurves with various amplitudes
and periods of 4~hours (a typical
KBO photometric variation period)
and carried out the
best-fit solution search described above.
We successfully recovered all
synthetic lightcurves with amplitudes
larger than around~15\%.
We can therefore place an upper limit
on possible lightcurve amplitudes
for \newbh ,
requiring that any such 
variation
must have an amplitude
less than 0.15~magnitudes
to be undetected by us.
The albedo variations and/or asphericity
of this body must be less than 15\%.

\subsection{\fv }
\label{fv}

\begin{figure}
\plotone{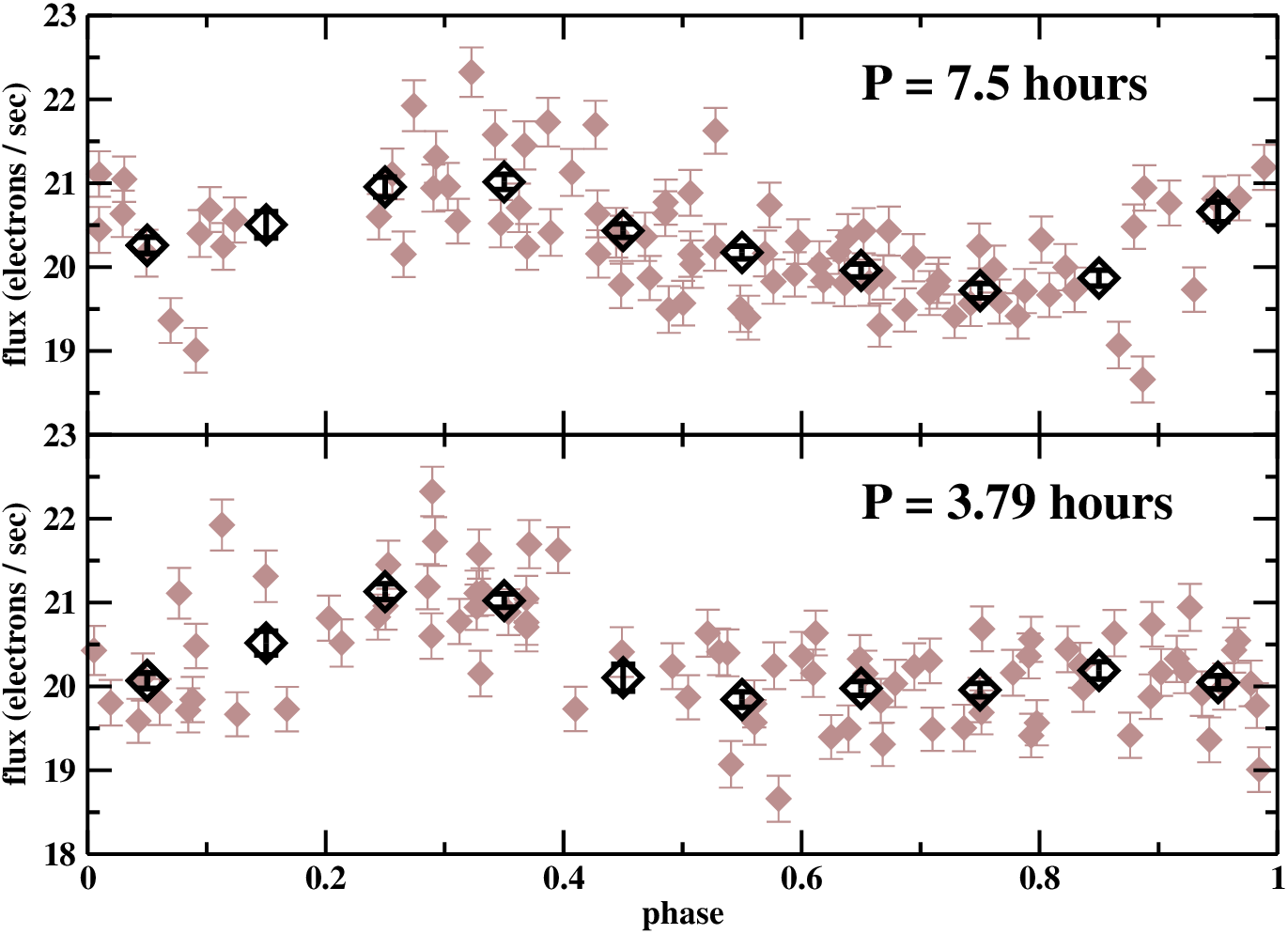}
\caption[]{
Observed data for \fv\ phased
at the best-fit period of 
7.5~hours (upper panel) and the second-best-fit
period of 3.79~hours (lower panel),
almost exactly half the best-fit period.
In both panels
the pale, filled diamonds show
the individual photometric measurements;
the large, open symbols show the
observed data binned
with bins 0.1~phase units wide.
The amplitude of the lightcurve
is identical (0.07~mag) for both
solutions.
The phased data in the top panel
appears double-peaked, with maxima
at phases of 0.3--0.35 and
0.95--1.0. These two peaks have
different shapes and are not 0.5~phase
units apart, so we conclude that this
lightcurve
is double-peaked and non-sinusoidal and that 7.5~hours is
the true rotation period of \fv.
However, we include the 3.79~hour
period in discussions below for completeness;
this lightcurve (lower panel) is also
significantly non-sinusoidal.}
\label{fvphase}
\end{figure}

The best-fit solution for
\fv\ gives a period of 7.5~hours.
There is a peak of nearly equal significance
at 3.79~hours, almost 
exactly half the best solution.
The amplitude of the lightcurve
is identical (0.07~mag) for both
solutions.
We compare the \fv\ observed data
phased at each of these two periods
in Figure~\ref{fvphase}.
The phased data in the top panel
appears double-peaked, with maxima
at phases of 0.3--0.35 and
0.95--1.0. These two peaks have
different shapes and are not 0.5~phase
units apart, so we conclude that this
lightcurve
is double-peaked and non-sinusoidal, and that 7.5~hours is
the true rotation period of \fv.
(We note a low-signal maximum for \fv\ in
Figure~\ref{period} at 
15~hours, which could be an alias of
the 7.5~hour rotation period.)
However, we include the 3.79~hour
period in discussions below for completeness;
this lightcurve (lower panel) is also
significantly non-sinusoidal. Arguably,
the true photometric period could
be 3.79~hours, with 7.5~hours an alias
of this true photometric period.
We discuss the implications of the
non-sinusoidal lightcurve below.

\section{Lightcurve modulations produced by surface features}
\label{surface}

Observations of KBOs are necessarily conducted with
the line of sight very close to the direction of illumination: in our
case, 1\fdg4 and 1\fdg7 for the new bodies and \fv, respectively.  In
this case we can ascribe lightcurve variations to some combination of
(1) variation in surface
composition and/or albedo that rotate through the observed
hemisphere;
(2) small-scale irregularities
(``facets'') that rotate through (un-)favorable orientations for
reflecting radiation to the observer;
or 
(3) changes  
in projected area of the rotating body due to its gross shape, often
approximated as an ellipsoid.

We will examine in turn these possible causes for the photometric
variations of
our measured KBOs, with the goal of extracting any possible
constraints on their internal structure or surface composition.  We
will focus primarily on the objects \newbf\ and \fv\
because their exceptionally
large and small photometric variations,
respectively, provide the
most interesting constraints.
The lightcurves of \newbg\ and \newbh\
are relatively unremarkable and do not help
differentiate among possible physical models,
so they will not be discussed further.

\subsection{Albedo effects}
\label{albedo}
If the large observed lightcurve variation of \newbf\ is due
entirely to albedo variations, surface patches with albedos that
differ by a factor of~2.5 are required
if the body has two distinct
hemispheres and the rotation pole is perpendicular to the line of
sight.  If either of these two assumptions is relaxed, the required
albedo range of the surface materials is even higher.

Little is known about KBO albedos. The few data points suggest a
range from a few percent to perhaps 20\% or 
more
\citep{alten01,jewitt01,groussin03,bt04,alten04,noll04b,stans04,cruik05,stans05}.
The canonical (but
unsupported; see \citet{alten04} and others) KBO albedo of 4\%, which
is 
based on comet albedos, is more than 4~times smaller than the
17\% albedo observed for (55565) 2002~AW$_{197}$ \citep{cruik05}.
This large albedo range could therefore plausibly exist on \newbf, 
though this possibility seems extreme and is not
consistent with the existing sparse KBO albedo data.
The range of albedos on Pluto's surface exceeds a factor
of~5 \citep{stern97}, although Pluto's atmosphere contributes
substantially to this effect and \newbf\ would not be expected to have
any atmosphere (because of its small size).
The two hemispheres of Iapetus have albedos differing
by a factor of~7, though this is likely due both
to being tidally locked to Saturn and to
contamination from other satellites.

These considerations favor shape over albedo as the primary cause of
the variability of \newbf, more so because the
main-belt asteroid (243)~Ida provides a (rocky) example of the shape
needed to produce this light curve (\S\ref{comparison}).

On the other hand, albedo and shape could
be correlated, as would be the case
with a large, fresh, bright crater.  Furthermore,
a crater or albedo feature could
potentially dominate the majority of a hemisphere of a
small body like \newbf , so we cannot
exclude the possibility of a wide range of surface reflectance on
\newbf.  
Additionally, KBO surfaces likely incorporate
volatiles that could potentially be mobilized
either through collisions or potentially
even seasonal thermal variations.
Migration of volatiles could plausibly create
patchy surfaces with albedo variations.

The small photometric variations for the other three bodies could easily be
explained by albedo variations on the surface.
We note that lightcurves from albedo variations need not be symmetric,
and the non-sinusoidal lightcurve of \fv\ suggests the
rotation of (two) bright spots past the subobserver point.
However,
it is unlikely that albedo
variations would conspire to reduce
the amplitude of an
otherwise large, shape-derived light curve
(this would require a dark long axis and
bright short axis of an elongated rotating body).
Therefore, the
possibility of albedo variation does not invalidate the
geophysical arguments
presented below.

\subsection{Facets on KBOs}
\label{facets}
A second possible explanation of the KBO lightcurves arises from study
of small bodies of the inner Solar System.  Complex shape models have
been determined for a number of asteroids and near
Earth objects (NEOs) not only from
spacecraft observations (e.g., \citet{thomas99,wilk02}) but also from
radar studies (see \citet{ast3} for a review).
Good
lightcurves have been measured for many asteroids and NEOs whose
shapes are known, and
several NEOs are observed to have
lightcurve amplitudes substantially larger or smaller
than their elongated shapes naively
would suggest \citep{pravec98,benner99,benner02}.

Because asteroids, KBOs, and
comets all are heavily
cratered bodies\footnote{For example,
recent imaging of
Comet Wild 2 by the Stardust mission
shows a very cratered comet
surface
\citep{wild2}.},
non-uniform facets (reflecting faces
and partially concave shapes) potentially can mask the true shape
of the body.
The
scenario in which faceted KBOs show lightcurves
larger than their gross shape would otherwise
suggest cannot be ruled out.
Thus, \newbf\ 
could have a
complicated topography that produces
a lightcurve that --- at least during
our observing season --- is substantially
larger than its gross shape might
otherwise indicate.  Conversely, the gross shape of \fv\ may be less
regular than its small amplitude
light curve suggests, a caveat to bear in mind for
the analyses below.
Facets on \fv\ could also produce 
the observed non-sinusoidal lightcurve.

\section{Geophysical considerations}
\label{geophysics}

In this section we regard the photometric variation as primarily a
result of the gross shape of the KBO, and examine the constraints on
internal strength and density that may be derived from the rotation
properties.  We will focus primarily
on the constraints imposed by the small 0.07~mag
amplitude of the \fv\ lightcurve.

\subsection{A simple shape model}
\label{shapes}

Observed KBO brightness variations
may be the result of the gross aspherical
shape of the body. A KBO may generally
be thought of as having three primary
axes, $a$, $b$, and $c$, where
$a\geq b\geq c$ and where rotation takes
place about $c$ in the 
minimized energy and angular momentum
state.
If this body is viewed 
equatorially, the ratio
$a/b$ determines the magnitude
of the observed lightcurve modulation
as $\Delta m = 2.5 \log (a/b)$.
\citet{ll03}
present a formalism for calculating
an observed magnitude variation
(for essentially  Lambertian bodies)
as a function of body shape and the
viewing angle $\theta$ between the rotation axis and the
(coincident) lines of sight and illumination (their
Equation~2).
The conditions in which amplitudes
smaller than 0.07~mag are produced
correspond to bodies of any shape
seen nearly pole-on ($\theta \approx 0$);
and
nearly spherical bodies
($a\approx b\approx c$)
seen at any angle.
Note that for a KBO in pole-on rotation, the coincidence of
illumination, line-of-sight, and rotation axes drives the lightcurve
amplitude to zero regardless of the body shape or surface properties.
In this configuration,
the low amplitude of the \fv\ lightcurve would
allow no definitive constraints on its
properties (although a useful constraint
can still be derived from the photometric period;
see below).
If \fv\ is significantly aspherical and
exactly pole-on at present, its lightcurve amplitude
should increase as it proceeds along its
250~year orbit; a 20~year observational
baseline could provide a pole-Earth
angle change of $\sim$30~degrees,
potentially revealing the equatorial
aspect and therefore shape of the body.
However,
\fv\ was targeted without regard to variability properties, so we can
consider its pole orientation to be a random variable.  Rotation axes
within 25\arcdeg\ of pole-on occur 
less than 10\% of the time for random
orientations, so we will exclude as unlikely any solution that requires 
$\theta<25\arcdeg$.
%
We note further
that, of the
10~KBOs and Centaurs with $7.5\leq H\leq 9.5$ (that is,
objects around the size of \fv)
and measured light curves, 
4~have
amplitudes $\le0.2$~mag (see Table~\ref{otherdata}).
Low variability is therefore
not rare in
the \fv\ size range,
consistent with the argument that
the small \fv\ lightcurve need
not be produced by an unlikely orientation.

If the small lightcurve is not produced by a nearly
pole-on orientation, several approaches lead to
interesting constraints on the physical
properties of \fv .
We consider in turn the possibilities that \fv\ has essentially zero
internal strength (a fluid); low internal strength (a rubble pile); or
material strength as in a monolithic
(consolidated) body.

\subsection{Fluid solutions}
\label{fluids}

\citet{chandra,hubbard}; and
\citet{tassoul} have discussed the
energy distributions and shapes 
of rotating, equilibrium, fluid
bodies, and we apply their
analyses here.
The physical state of a rotating, fluid (strengthless)
body depends on the angular momentum
and distribution of matter. 
Non- or slowly-rotating fluid bodies are
generally spherical. Moderate rotation
produces a Maclaurin spheroid in
which $a=b\gtrsim c$,
and faster rotation
results in a triaxial Jacobian ellipsoid
in which $a>b\gtrsim c$.
Hubbard
defines the dimensionless
rotation rate $\Omega$ as

\begin{equation}
\Omega^2\equiv\frac{\omega^2}{2\pi G \rho}
\label{hubbeqn}
\end{equation}

\noindent where 
$\omega$ is the angular
rotation rate and 
$\rho$ is the bulk density
of the body;
in this formalism,
the transition from Maclaurin to
Jacobian bodies occurs at the 
bifurcation point
$\Omega^2=0.19$.

The maximum value for the
dimensionless rotation rate ($\Omega$)
is reached at the bifurcation
point.
Thus,
the minimum density for a fluid \fv\ is 0.67~g~cm$^{-3}$
for a rotation period of 7.5~hours.
All densities greater than this produce
two theoretically viable solutions,
one representing the Jacobian ellipsoid
branch of solutions and one the
the Maclaurin spheroid 
branch.
Here we consider each branch in turn.

\subsubsection{Jacobian ellipsoid solution}

\begin{figure}
\plotone{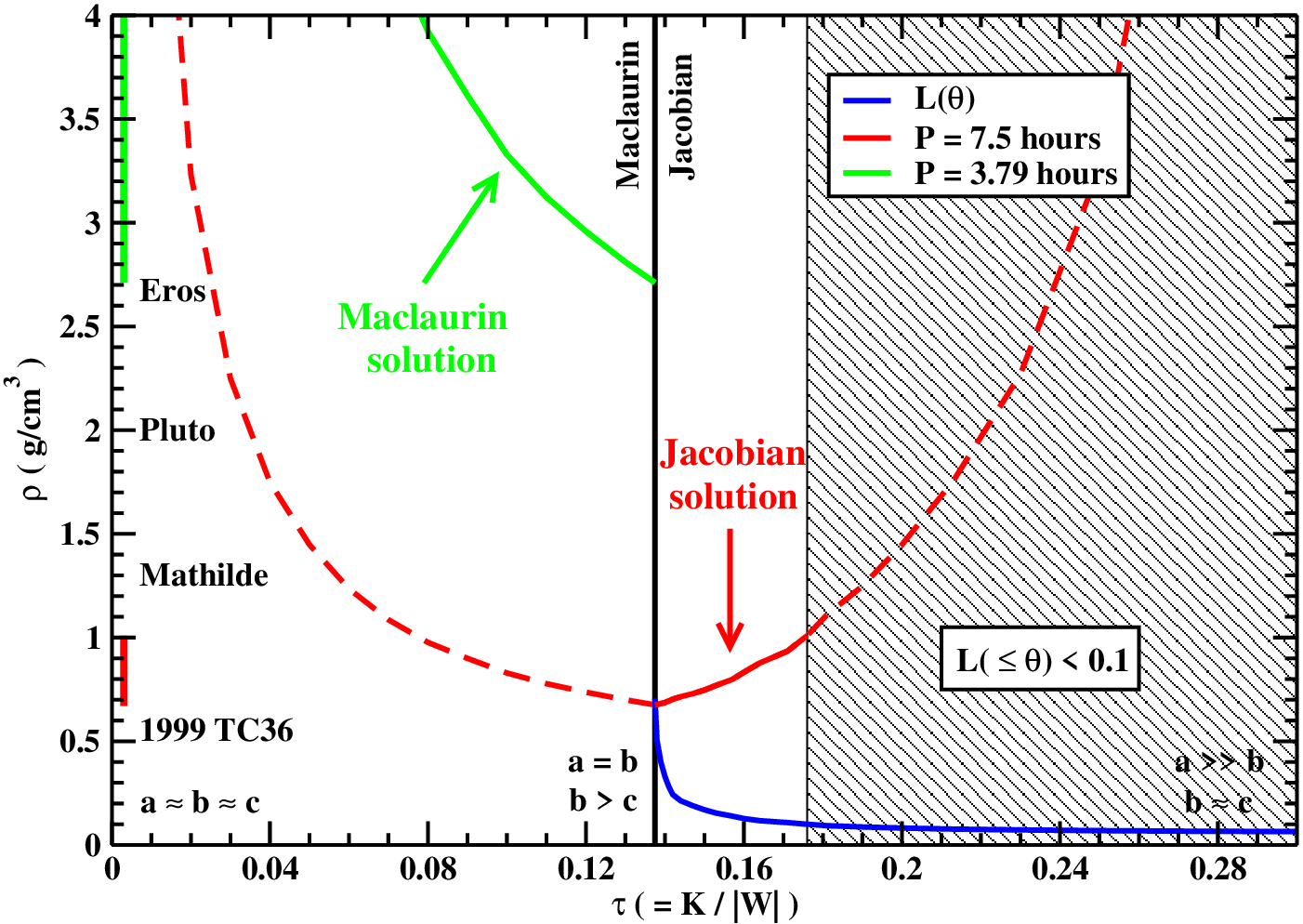}
\caption[]{{\tiny Density
versus energy
state
for geophysical equilibrium
fluid solutions for
\fv, where
$\tau$ is 
the ratio of rotational
kinetic energy (K) to
the absolute value of the
gravitational potential energy (W).
The thick vertical
line at $\tau=0.1375$ marks the 
bifurcation point between 
Maclaurian solutions and
Jacobian solutions.
Red curves show solutions for rotation
a rotation period of 7.5~hours.
The green curve shows the solution
for a 3.79~hour period, possible
in the Maclaurin branch.
The blue
curve shows $L(\leq\theta)$, 
the probability of the orientation
required for a given $\tau$
(this probability is plotted between
zero and one using the labels on
the vertical axis). The grey hatched
region indicates $L(\leq\theta ) < 0.1$,
which we consider unlikely.
The only likely Jacobian solutions
(solid red line)
have densities 0.67--1.0~g~cm$^{-3}$
(as indicated by the red bar along
the vertical axis).
The right dashed red curve shows
unlikely Jacobian solutions (low
geometric probability).
The left dashed red curve shows
solutions
that are unlikely
because 
a double-peaked lightcurve is an unlikely
to reflect the true rotational period of
a Maclaurin body.
Instead,
the green curve 
shows the possible
densities for Maclaurin solutions
(indicated by the green bar along
the vertical axis).
The densities of asteroids Mathilde and 
Eros 
\citep{yeomans97,yeomans99,veverka99},
Pluto \citep{olkin03},
and
KBO
(47171) 1999~TC$_{36}$
\citep{stans05}
are indicated.
The approximate relationships 
among $a$, $b$, and $c$ are 
given to lend intuition to
the solution shapes described.
%
%
%
%
}}
\label{fvfig}
\end{figure}

%
If we assume that the observed \fv\
lightcurve is derived from the gross
shape of the body, then we require
the branch of solutions corresponding
to a triaxial
Jacobian ellipsoid
where $a>b$.
The rotation period must be 7.5~hours:
if the best-fit solution of 7.5~hours is used,
its double-peaked nature implies that it
is a complete rotation period, whereas
if the second-best-fit solution 
of 3.79~hours is used, its single
peak implies that 3.79~hours corresponds
to only a half period (since the lightcurve
is shaped-derived for a Jacobian body).
Hence, we know $\omega$ (the angular rotation
rate).
Tassoul introduces $\tau$,
which describes the energy state
of a rotating body and
which is the ratio of 
rotational
kinetic energy (K) to
the absolute value of the
gravitational potential energy (W):
$\tau=K/|W|$, where $\tau$ is small
for nearly spherical bodies and
increases for bodies with increasing
asphericities (Figure~\ref{fvfig}).
Tassoul 
shows the relationship 
between $\Omega^2$ and $\tau$;
from Equation~\ref{hubbeqn} and 
our knowledge of $\omega$, we convert
this relationship to 
$\rho$ as a function of $\tau$.
This result --- bulk density
as a function of the energy state of
the body for a Jacobian ellipsoid --- is shown in
Figure~\ref{fvfig} as the red line on the 
right half of the plot.

\citet{chandra} tabulates the relationship
between $b/a$, $c/a$ and
$\omega^2/(\pi G \rho)$
(what Chandrasekhar writes as $\Omega$
we write here as $\omega$, the angular
rotation rate).
We therefore can derive the relationship
between $b/a$, $c/a$, and $\rho$,
and consequently between
$b/a$, $c/a$, and $\tau$.
From $b/a$ and the observed
$\Delta m$, we calculate
the required viewing angle
$\theta$ following Equation~2 of \citet{ll03},
deriving $\theta$ as a function
of $\tau$. Lastly,
we introduce $L(\leq\theta )=(1-\cos\theta)$,
which 
is the probability, from simple geometric
arguments, that a randomly oriented
rotation pole has an orientation angle
less
than or equal to $\theta$.
We show $L(\leq\theta)$
as a function of $\tau$
in Figure~\ref{fvfig}.

We exclude $L(\leq\theta)<0.1$ solutions
as improbable.
For \fv ,
$L(\leq\theta )\geq 0.1$ corresponds to
$a/b=2.8$ and $a/c=3.3$ and
to
$\rho\leq 1.0$~g~cm$^{-3}$ (Figure~\ref{fvfig}).
Thus, for Jacobian solutions --- the {\em only} fluid solutions
in which the photometric lightcurve is derived
from the gross aspherical shape of the body --- the bulk density of \fv\
must be 
0.67--1.0~g~cm$^{-3}$ (red solid line
in Figure~\ref{fvfig}).

By comparison,
assigning Pluto's density of 2~g~cm$^{-3}$
\citep{tb97}
to \fv\ and assuming a Jacobian
solution, we find
$a/b\approx 5.6$
and $a/c\approx 6.1$, a long, thin body
whose shape would be the most extreme in 
the Solar System:
asteroid 216~Kleopatra (the ``dogbone
asteroid'') has
$a/b$ around~2.3 
\citep{ostro00,hest02}.

\subsubsection{Maclaurin spheroid solutions}

The Maclaurin branch
of solutions represents oblate spheroids with
$a=b$ and rotation about $c$.
In this scenario,
there is no 
modulation of cross-sectional area during rotation.
Here,
photometric variations
must be due to surface features,
either albedo variations (Section~\ref{albedo})
or facets (Section~\ref{facets}).
The rotation period for \fv\ could
therefore be the second-best-fit
solution of 3.79~hours, as a double-peaked
lightcurve (i.e., the 7.5~hour
period) would have to be produced by 
the unlikely configuration of bright (or dark)
spots on opposite hemispheres of a body.
We show the solution for a 3.79~hour
rotation period in green in Figure~\ref{fvfig};
the 
minimum allowable density is
2.7~g~cm$^{-3}$, slightly higher than
that of the (presumed rocky) NEO
433~Eros.
Though there is formally no maximum,
densities larger than 4~g~cm$^{-3}$
are remarkably unlikely.

If \fv\ is a Maclaurin body
with a double-peaked lightcurve
and rotation period of 7.5~hours -- a situation we
consider unlikely because of the requirement
that like surface features be antipodal,
a seemingly improbable configuration --
then the red dashed
line on the left half of Figure~\ref{fvfig}
obtains.


\subsubsection{Summary of fluid solutions}

To summarize,
if \fv\ is strengthless (fluid),
there are two primary solutions, both of which would be surprising: 
(1) \fv\ is a triaxial body with
$\rho$ in the range 0.67--1.0~g~cm$^{-3}$, implying a very high ice
fraction or very high porosity; or
(2) \fv\ is an oblate spheroid with a ``bright spot''
(or else in an unlikely pole-on orientation)
and has density $\rho\ge 2.7$~g~cm$^{-3}$, which
is
quite high for a body that is expected
to be rock and ice with moderate porosity.
We are forced to conclude that a strengthless \fv\ must have a composition that
is either surprisingly ice-rich or surprisingly rock-rich, implying
that it is a fragment of a differentiated body.
However, the (implied) sphericity and relatively
large size of \fv\
do not favor the fragment interpretation.

Geophysical fluid solutions for
\newbf\ place no surprising constraints on
its density:
%
for Jacobian solutions, density is in the expected
range 0.5--2.5~g~cm$^{-3}$, and for the
``Maclaurian-with-a-spot'' model, the lower limit
the density is 0.5~g~cm$^{-3}$.

\subsection{KBOs with non-zero internal friction}
\label{strength}

In the previous section, we found that
the fluid solutions for \fv\
require surprising densities (or may 
be geometrically unlikely).
However, if KBOs are rubble piles
made of rocks and ice,
we can relax the fluid assumption by allowing
these rotating bodies to have intrinsic strength.

\citet{hols01} has studied 
the effects on body shapes
and rotation rates of allowing
cohesionless rubble piles to
experience internal friction
akin to
the strength exhibited
by a pile of sand.
The results are described in terms
of $\phi$, the angle of internal
friction (or angle of repose).
Fluids necessarily have $\phi=0^\circ$;
typical terrestrial soils have
$\phi\lesssim 30^\circ$.
Allowing bodies to have non-zero
internal friction allows shapes
that depart from the 
Maclaurin-Jacobian spheroid/ellipsoid
sequence.
%
%
We must assume a density to
constrain internal friction;
we first consider \fv\ and assume
$\rho=1$~g~cm$^{-3}$.

\begin{figure}
\plotone{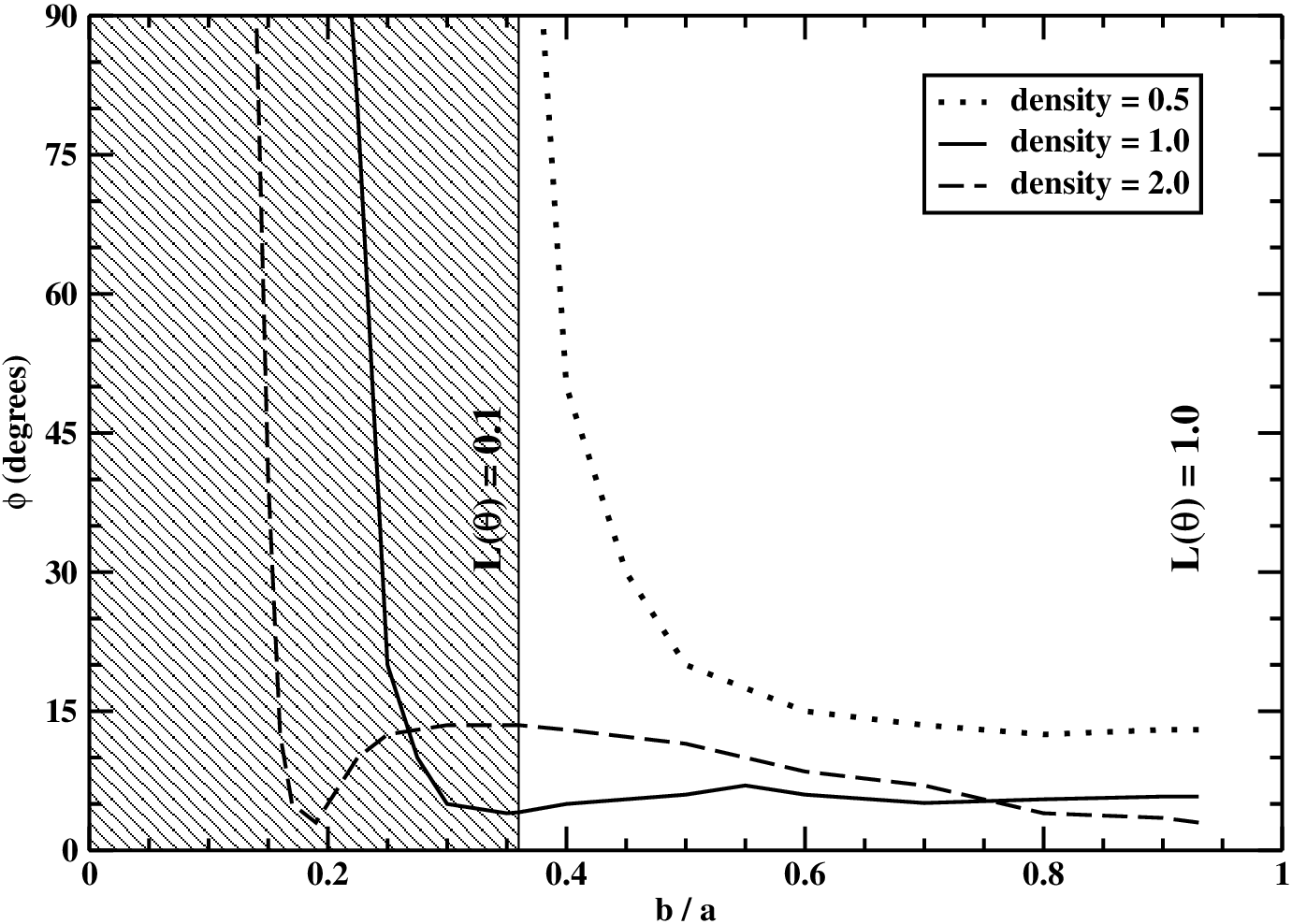}
\caption[]{Angle of internal
friction $\phi$ as a function
of axis ratio $b/a$ for 
\fv , for three different
densities (as shown).
These curves are slices through
a complicated, undulating surface
(see Holsapple Figure~3).
Each value of $b/a$ implies 
a specific viewing angle
$\theta$, as constrained by
the observed lightcurve amplitude;
$L(\leq\theta ) < 0.1$ for
$b/a < 0.36$ (hatched region).
For densities of 1~or 2~g~cm$^{-3}$,
$\phi$ is less than around
15\arcdeg\ throughout the region
of probable orientation.
However,
$\rho=0.5$~g~cm$^{-3}$ requires solutions
with large $\phi$ for probable
orientations.
Therefore, densities much smaller than
1~g~cm$^{-3}$ may be unlikely,
but densities of 1--2~g~cm$^{-3}$
provide solutions with reasonable
angles of internal friction.
}
\label{phi}
\end{figure}

We consider the prolate
spheroid case
$a>b=c$ (e.g., Holsapple Figures~3 and~5),
assumed to be rotating about $c$
with a rotation period of 7.5~hours.
Holsapple assumes equatorial viewing,
which we are not restricted to in our
analysis.
Instead we may consider a range of
axis ratios, where each value of
$b/a$ implies a specific viewing
angle $\theta$ as constrained by
the observed lightcurve amplitude.
Holsapple's
dimensionless rotation rate
(which he defines as
$\Omega=\omega/\sqrt{\rho G}$)
is determined by our assumption
of $\rho$ and knowledge of
$\omega$.
If we require $L(\leq\theta )\geq 0.10$,
then $b/a$ for \fv\ must be
in the range~0.36--0.93.
Throughout this range, 
$\phi \approx 5^\circ$ (Figure~\ref{phi}).
%

We now relax our density assumption.
Carrying out the same analysis for
$\rho=2.0$~g~cm$^{-3}$,
we find that
$\phi$ is less than
15$^\circ$ for all orientations
with $L(\leq\theta ) \geq 0.10$
(Figure~\ref{phi}).
However, for
$\rho=0.5$~g~cm$^{-3}$,
the minimum $\phi$ is
around 13$^\circ$ and
probable orientations require
large $\phi$.

We thus find that the internal friction 
for \fv\ can reasonably be small but non-zero.
Additionally, densities much less
than 1~g~cm$^{-3}$
have solutions
that increasingly require $\phi > 30^\circ$,
an unlikely physical scenario.
Therefore, the physical picture that 
emerges is the following:
\fv\ can readily be a rubble pile 
with density 1--2~g~cm$^{-3}$
and small angles
of internal friction.
This solution does not require
excessive porosity (from density
estimates).
This weak rubble pile body --- multiply
impacted into a collection of blocks that has only
small internal friction --  may be
nearly, but not quite, relaxed to
geophysical fluid equilibrium configuration.
This interpretation also allows for
the non-sinusoidal lightcurve observed
for \fv\ is the body may still
contain blocks that are out of 
fluid equilibrium.


For \newbf ,
$\rho=0.5$~g~cm$^{-3}$ produces solutions
for $\phi$ of generally less than 15~degrees.
\newbf\ solutions for 
$\rho=1.0$~and 2.0~g~cm$^{-3}$ increasingly
include
$\phi$ of 18--25~degrees
as well as narrow regions
(in $b/a$)
of acceptably low internal
friction and
very high internal
friction (see Holsapple Figure~3).
We conclude that no combination of
density and internal friction are
precluded for \newbf ,
though $\phi$ of 25~degrees is 
larger than is observed for most asteroids \citep{hols01}.
This may indicate that \newbf\ is more likely to be kept out of
equilibrium by monolithic
strength (see below) than by rubble-pile friction.

\subsection{Material Strength}

We considered above small-grained
rubble piles,
gravitational aggregates of 
loose material
(see, e.g.,
\citet{lein00}).
Rubble pile
bodies can also have ``strength''
if they consist of some large 
blocks in a mixture of smaller
rubble (D.\ Richardson, pers.\ comm.;
\citet{dcr05}).
Finally,
small bodies in the Solar System can
be strong if they are monolithic
bodies -- essentially a single chip
or fragment from a larger body, or the frozen remnant of a previously
fluid body.
In theory, such bodies could be
either rock or ice.

\fv\ may be far from an equilibrium rotation configuration if it is
composed of one (or a few) blocks of solid material.
Surface gravity scales linearly with the product
of size and density; using this rough guideline,
topography on \fv\ on the order of 10~km
(enough to create 10\% asymmetries)
could
easily be supported by
strength (E.\ Apshaug, pers.\ comm.).
If monolithic strength exists on \fv\ and
supports topography, though, then 
the question exists why \fv\ is nearly
symmetric as its lightcurve suggests.
Additionally,
dynamical
arguments (see below) suggest a much-impacted
and therefore completely shattered
and rubble-like body.
Irregular surface topography -- that need
not be supported by monolithic strength
and is similar to
the facets discussed above -- could produce
the non-sinusoidal lightcurve observed
for \fv.

The pressure at the center of
a planetary body may be approximated
as 
$GM^2/r^4$ where
$M$ and $r$ are the mass and
radius of the body.
When $b/a$ for \newbf\ is
in the range 0.3--0.9, the overburden
pressure produced by the asymmetric
shape
is 
1--10~kilobars.
The strength of clean laboratory
ice is approximately 10~kilobars
and that of snow is 0.01--0.1~kilobars (E.\ Asphaug, pers.\ comm.).
Thus, the central stress in
\newbf\ could easily be supported
by its material strength.
\newbf\
could
easily be a rotating, coherent
monolith with a substantially
aspherical shape.


\section{Discussion}
\label{discussion}

\subsection{Summary of Constraints}

\fv\ is a modest-sized KBO of diameter 116~km if the albedo is 0.10.
There is a small chance that the low amplitude of the \fv\ light curve
is attributable to pole-on rotation, but otherwise it must be a
remarkably spherical body.
Topography is allowed by strength arguments,
but the size of \fv\ suggests that it should
have been impacted many times
(\citet{durda00}; see below)
and hence be a rubble pile, not
a monolithic body.
The solutions for a fluid body
require either a surprisingly low ($<$1~g~cm$^{-3}$) or high
($>$2.7~g~cm$^{-3}$) density, likely requiring \fv\ to be a remnant of
a differentiated body.
The former solution may imply a remarkably
large porosity.

A more plausible solution is that \fv\ is a
rubble pile of density 1--2~g~cm$^{-3}$, held slightly out of the
minimum-energy shape by internal friction among constituent blocks
that are relatively small.
The non-sinusoidal light curve of \fv\
requires surface inhomogeneity or a departure from ellipsoidal shape,
but either effect need only be slight,
and the latter is easily allowed
for by a nearly-relaxed rotating body with 
non-zero internal friction.

The flux from the small body \newbf\
(20~kilometers diameter, for albedo of 
10\%) varies by a factor~$>$2.5 over the light curve.
Such large-amplitude variation is achievable if the body is an
irregularly-shaped collisional remnant consisting of one or a small
number of coherent fragments supported by material strength. 
Alternately, extreme albedo variations
would be required to
explain the 1.09~mag lightcurve variation,
perhaps with one impact-generated clean
ice hemisphere contrasting with a darker
(5\%--10\% albedo, consistent with that
measured for other KBOs and Centaurs) hemisphere.

The ACS data for
\newbg\ and \newbh\
do not allow placing any interesting
constraints on surface or internal composition.

\subsection{Collisions in the Kuiper Belt}
\label{collisions}

The Kuiper Belt is generally thought of
as a collisionally evolved population.
This environment can readily produce
facets on KBOs; impacts likely can also
produce albedo features on KBOs
through cratering; and
elongated objects can be produced through
fragmentation.
However, 
the nearly spherical \fv\ must also
be created through, or survive, collisional evolution.

\citet{durda00} have calculated
the timescale for disruptive collisions
based on the present environment in
the Kuiper Belt and assuming
the pre-\citet{gmb} understanding
(i.e., overestimation)
of the small-end size distribution.
They found that the timescale for
disrupting a 100~km KBO is substantially
longer than the age of the Solar 
System.
Thus, \fv\ is likely not a fragment
that was recently created.
Instead, 
the size of \fv\ likely records
the timescale and efficiency of
accretion in the Kuiper Belt: \fv\
represents an intermediate product
of the accretion process that formed
Kuiper Belt giants like Quaoar and Pluto.
\citet{lein00} showed 
that pairwise accretion of 
rubble piles can produce
both spherical and
aspherical bodies.
Thus, both \fv\ and 
Quaoar, which
potentially has a 10\% 
asphericity as indicated by
its lightcurve \citep{ortiz03b},
can have gross shapes that are
the direct result of 
rubble
pile accretion.
Additionally, \fv\
may have been impacted
many times since its
formation, resulting in a completely
shattered body (consistent with \citet{pan05}); we note that early
in the Solar System's history, the space
density of bodies in the Kuiper Belt was higher 
than today
and the impact rate was
higher than at present.
Multiple collisions can produce
the small internal friction values
we derived in Section~\ref{strength}.
A consistent
picture for \fv\ is
therefore
that of a body that accreted to approximately
its present size; has been substantially
shattered due to extensive collisions;
has little internal friction due to 
its rubble pile nature; and is nearly,
but not completely, relaxed, thus
nearly attaining a rotating
fluid equilibrium
state.


\citet{durda00} find that the 
disruption timescale for a 
30~kilometer body is also longer
than the Solar System, implying
that formally a 30-km KBO would reflect
primordial growth, not collisional disruption.
Including the \citet{gmb} results
will increase the disruption timescale
for bodies of this size
because of the dearth of small bodies.
Thus, the picture for \newbf\ may
be somewhat complicated, as an elongated
body is implied by its lightcurve
amplitude.
\newbf\ may be a fragment from an
unusual, but not wildly improbable,
collision between 50--100~kilometer bodies.
Furthermore, this collision could
have occured billions of years ago
when the space density of KBOs
was higher,
before later dynamical sculpting
and mass loss (e.g., \citet{morby03,gomes05}).
Our interpretation of 
the \newbf\ data is
that the body is an elongated KBO
(though not necessarily a monolithic
body),
and the collisional fragment solution is
appealing in this case.

Alternately, \newbf\ may have a complicated
surface that produces a lightcurve larger
than its gross shape would suggest.
Eons of impacts certainly could produce
arbitrarily complicated surface 
topographies,
though
\citet{kory03} show that cumulative
small impacts on rotating asteroids
tend to lead to oblate shapes,
which cannot produce the observed
lightcurve.
Understanding this object, in
the absence of many comparably small
KBOs, requires us to look elsewhere
in the Solar System (Section~\ref{comparison}).

%


\subsection{Comparison to other KBOs}
\label{densities}

We list in Table~\ref{otherdata}
the 65~KBOs and Centaurs (excluding comets) for which
lightcurve measurements or upper 
limits have been published;
37~of these have reported
periodic lightcurve
amplitudes, typically greater than
$\sim$0.1~magnitudes.
Most of these bodies have
implied rotational periods
(or half-periods for double-peaked lightcurves)
in the range 3--10~hours, similar to the
periods derived for our HST/ACS KBO
observations.
Note that
these surveys certainly do
not represent a complete nor
random sample: some
non-detections are likely
unreported, and these observations
represent mostly the brightest (largest)
KBOs, so biases certainly exist
in this compiled literature sample.
Nevertheless, interesting results
can be derived.

The amplitude we derive for
\newbf,
together with the recently
measured amplitude of
1.14~mag for 2001~QG$_{298}$ \citep{sj04},
are 
the largest amplitude
variations (to date) for KBOs and
Centaurs.
Additionally, our data show
lightcurves for the faintest
(and therefore smallest) KBOs,
to date. However,
neither the large lightcurve amplitude
of \newbf\ nor the fact that the small
bodies
\newbf\ and \newbg\
have lightcurves
are particularly remarkable in the Solar System,
as many
small asteroids are known to have lightcurve
variations larger than
1~magnitude, including
some kilometer-sized NEOs
\citep{pravec02}.

\begin{deluxetable}{lcccccc}
\tablewidth{0pt}
\tablecaption{KBO and Centaur lightcurve data}
\tablehead{
\colhead{Desig.} & 
\colhead{Number} &
\colhead{Name} &
\colhead{Period(s)\tablenotemark{a}} & 
\colhead{Amplitude(s)} &
\colhead{H\tablenotemark{b}} & 
\colhead{References} \\
\colhead{} &
\colhead{} & 
\colhead{} &
\colhead{(hours)} &
\colhead{(mag)} &
\colhead{}}
\startdata
\nodata		& \nodata & Pluto & 153.6\tablenotemark{c} & 0.33 & -1 & 1 \\
2003~EL$_{61}$  & \nodata & \nodata & 3.9 & 0.28 & 0.1 & 2 \\
\nodata         & \nodata & Charon & 153.6\tablenotemark{c} & 0.08 & 1 & 1 \\
2003~VB$_{12}$  & 90377   & Sedna  & 10.3                  & 0.02 & 1.6 & 3 \\
2002~LM$_{60}$  & 50000   & Quaoar  & 17.7\tablenotemark{d} & 0.13 & 2.6 & 4 \\
2001~KX$_{76}$     & 28978   & Ixion   & \nodata & $<$0.05 & 3.2 & 5 \\
2002~TX$_{300}$    & 55636   & \nodata & 7.89, 8.12, 12.10  & 0.08, 0.09  & 3.3 & 5,6 \\
2002~UX$_{25}$     & 55637   & \nodata & \nodata & $<$0.06 & 3.6 & 5 \\
                   &         &         & 14.4\tablenotemark{d}, 16.8\tablenotemark{d} & 0.2  & 3.6 & 7 \\
2000~WR$_{106}$ & 20000 & Varuna & 6.34\tablenotemark{d} & 0.42 & 3.7 & 8,9 \\
2003~AZ$_{84}$     & \nodata & \nodata & 6.72    & 0.14    & 4.0 & 5 \\
2001~UR$_{163}$    & 42301   & \nodata & \nodata & $<$0.08 & 4.2 & 5 \\
1996~TO$_{66}$     & 19308   & \nodata & 3.96    & 0.26 & 4.5   & 5 \\
                   & & & 6.25\tablenotemark{d} & 0.12, 0.33 & 4.5 & 10 \\
1999~DE$_{9}$      & 26375   & \nodata & \nodata & $<$0.10 & 4.7 & 9 \\
2000~EB$_{173}$    & 38628   & Huya    & \nodata & $<$0.06 & 4.7 & 9 \\
                   &         &         & 6.75    & $<$0.1  & 4.7 & 11 \\ 
2001~QF$_{298}$    & \nodata & \nodata & \nodata & $<$0.12 & 4.7 & 5 \\
1995~SM$_{55}$     & 24835   & \nodata & 4.04    & 0.19  & 4.8  & 5 \\
1998~WH$_{24}$     & 19521   & Chaos   & \nodata & $<$0.10 & 4.9 & 9 \\
1999~TC$_{36}$     & 47171   & \nodata & \nodata & $<$0.06 & 4.9 & 5 \\
2000~YW$_{134}$    & 82075   & \nodata & \nodata & $<$0.1  & 5.1 & 5 \\
1996~GQ$_{21}$     & 26181   & \nodata & \nodata & $<$0.10 & 5.2 & 9 \\
1997~CS$_{29}$     & 79360   & \nodata & \nodata & $<$0.08 & 5.2 & 9 \\
2002~VE$_{95}$     & 55638   & \nodata & \nodata & $<$0.06 & 5.3 & 5 \\
1996~TL$_{66}$     & 15874   & \nodata & \nodata & $<$0.06 & 5.4 & 12 \\
2001~CZ$_{31}$     & \nodata & \nodata & \nodata & $<$0.20 & 5.4 & 9 \\
2001~QT$_{297}$    & 88611   & \nodata & \nodata & $<$0.15 & 5.5 & 13 \\
2001~KD$_{77}$     & \nodata & \nodata & \nodata & $<$0.07 & 5.6 & 5 \\
1998~SM$_{165}$    & 26308   & \nodata & 4.00    & 0.56    & 5.8 & 14 \\
1998~SN$_{165}$    & 35671   & \nodata & 5.03    & 0.15    & 5.8 & 15 \\
1999~KR$_{16}$     & 40314   & \nodata & 5.93, 5.84 & 0.18 & 5.8 & 9 \\
2000~GN$_{171}$ & 47932   & \nodata & 8.33\tablenotemark{d} & 0.61 & 6.0 & 9 \\
1998~XY$_{95}$     & \nodata & \nodata & \nodata & $<$0.1  & 6.2 & 16 \\
2001~FP$_{185}$    & 82158   & \nodata & \nodata & $<$0.06 & 6.2 & 5 \\
2001~FZ$_{173}$    & 82155   & \nodata & \nodata & $<$0.06 & 6.2 & 9 \\
2001~QG$_{298}$    & \nodata & \nodata & 6.89    & 1.14    & 6.3 & 17 \\
2001~QT$_{297}$B\tablenotemark{e}   & 88611B  & \nodata & 4.75    & 0.6     & $\sim$6.3 & 13 \\
1996~TS$_{66}$     & \nodata & \nodata & \nodata & $<$0.16 & 6.4 & 12 \\
1997~CU$_{26}$     & 10199   & Chariklo & \nodata & $<$0.1 & 6.4 & 18 \\
1977~UB  & 2060    & Chiron  & 5.92\tablenotemark{d} & 0.09 & 6.5 & 19 \\
1998~VG$_{44}$     & 33340   & \nodata & \nodata & $<$0.10 &  6.5 & 9 \\
1996~TP$_{66}$     & 15875   & \nodata & \nodata & $<$0.04 & 6.8 & 20 \\
1993~SC            & 15789   & \nodata & \nodata & $<$0.04 & 6.9 & 12 \\
1992~AD            & 5145    & Pholus  & 9.98    & 0.15--0.60\tablenotemark{f}    & 7.0 & 21,22,23 \\
1994~VK$_{8}$      & 19255   & \nodata & 3.9, 4.3, 4.7, 5.2 & 0.42 & 7.0 & 12 \\
1996~TQ$_{66}$     & \nodata & \nodata & \nodata & $<$0.22 & 7.0 & 12 \\
1994~TB            & 15820   & \nodata & 3.0, 3.5 & 0.26, 0.34 & 7.1 & 12 \\
1998~BU$_{48}$     & 33128   & \nodata & 4.9, 6.3 & 0.68 & 7.2 & 9 \\
2002~CR$_{46}$     & 42355   & \nodata & \nodata & $<$0.05 & 7.2 & 5 \\
1997~CV$_{29}$     & \nodata & \nodata & 16      & 0.4 & 7.4 & 24 \\
1995~QY$_{9}$      & 32929   & \nodata & 3.5     & 0.60  & 7.5  & 12 \\
1998~HK$_{151}$    & 91133   & \nodata & \nodata & $<$0.15 & 7.6 & 9 \\
2000~QC$_{243}$    & 54598   & Bienor & 4.57    & 0.75    & 7.6 & 11 \\
1995~DW$_{2}$      & 10370   & Hylonome & \nodata & $<$0.04 & 8.0 & 12 \\
{\bf \fv }                & {\bf \nodata } & {\bf \nodata } & {\bf 7.5\tablenotemark{d} }    & {\bf 0.07  }    & {\bf 8.2 }  & {\bf this work} \\
2002~PN$_{34}$     & 73480   & \nodata & 4.23, 5.11 & 0.18 & 8.2 & 11 \\
1999~TD$_{10}$ &29981 &\nodata &15.45\tablenotemark{d} & 0.65& 8.8 & 11,25,26 \\
1995~GO            & 8405    & Asbolus & 4.47    & 0.55 & 9.0    & 18,27 \\
2001~PT$_{13}$ & 32532 & Thereus & 8.3\tablenotemark{d} & 0.16 & 9.0 & 11,28 \\
2002~GO$_{9}$      & 83982   & \nodata & 6.97, 9.67 & 0.14 & 9.3 & 11 \\
2000~EC$_{98}$     & 60558   & \nodata & 26.8\tablenotemark{d}    & 0.24   & 9.5 & 7 \\
1993~HA$_{2}$      & 7066    & Nessus  & \nodata & $<$0.2  & 9.6 & 18 \\
1999~UG$_{5}$      & 31824   & Elatus  & 13.25   & 0.24  &   10.1 & 29 \\
{\bf \newbg  }           & {\bf \nodata }  & {\bf \nodata }  & {\bf 4.2 }    & {\bf 0.18 }    & {\bf 10.7 } & {\bf this work} \\
1998~SG$_{35}$	   & 52872   & Okyrhoe & 16.6\tablenotemark{d}    & 0.2   & 11.3 & 30 \\
{\bf \newbf      }       & {\bf \nodata } & {\bf \nodata } & {\bf 9.1, 7.3 } & {\bf 1.09  } & {\bf 11.7 } &  {\bf this work} \\
{\bf \newbh     }        & {\bf \nodata } & {\bf \nodata } & {\bf \nodata } & {\bf $<$0.15 } & {\bf 11.9 } & {\bf this work} \\
\enddata
\tablenotetext{a}{Photometric periods except for
double-peaked lightcurves, where the (presumed)
rotation period is listed.}
\tablenotetext{b}{Absolute magnitude: the (hypothetical) magnitude
the object would have at zero phase angle and
geocentric and heliocentric distances of 1~AU. Values from the
Minor Planet Center database.}
\tablenotetext{c}{Tidally locked}
\tablenotetext{d}{Double-peaked lightcurve}
\tablenotetext{e}{``2001~QT$_{297}$B'' is the binary
companion to 2001~QT$_{297}$.}
\tablenotetext{f}{Changes in Pholus' lightcurve
amplitude over the past decade can be explained by
changing viewing geometry to that Centaur
\citep{tegler05}.}
\tablerefs{(1) \citet{buie97};
(2) \citet{rabinowitz05};
(3) \citet{gaudi};
(4) \citet{ortiz03b};
(5) \citet{sj03};
(6) \citet{ortiz04}; 
(7) \citet{rousselot05};
(8) \citet{js};
(9) \citet{sj};
(10) \citet{hainaut};
(11) \citet{ortiz03a};
(12) \citet{rt99};
(13) \citet{osip03}
(14) \citet{rom01};
(15) \citet{peix02};
(16) \citet{cb01};
(17) \citet{sj04}
(18) \citet{davies98};
(19) \citet{bus89};
(20) \citet{cb99};
(21) \citet{buie92};
(22) \citet{farnham01};
(23) \citet{tegler05};
(24) \citet{chorney};
(25) \citet{rous03};
(26) \citet{choi03}:
(27) \citet{kern00};
(28) \citet{farnham03};
(29) \citet{gut};
(30) \cite{bauer03}}
\tablecomments{Multiple measurements have been made
for several bodies. 
For non-detections, we cite here only the most
sensitive measurement. For lightcurve determinations
that are in agreement with each other, we cite
all references
on the same line. For lightcurve determinations
that disagree where there is not clearly a superior
measurement, we list the conflicting results on
separate lines. Data reported here for the first
time are shown in bold.}
\label{otherdata}
\end{deluxetable}

\citet{ph00} derive a simple expression
that approximates the critical (minimum)
period ($P_c$, in hours) 
for a rotating body as a function
of density
and lightcurve amplitude
(in magnitudes):

\begin{equation}
P_c \approx 3.3 \sqrt{\frac{1+\Delta m}{\rho}}.
\label{pcrit}
\end{equation}

\begin{figure}
\plotone{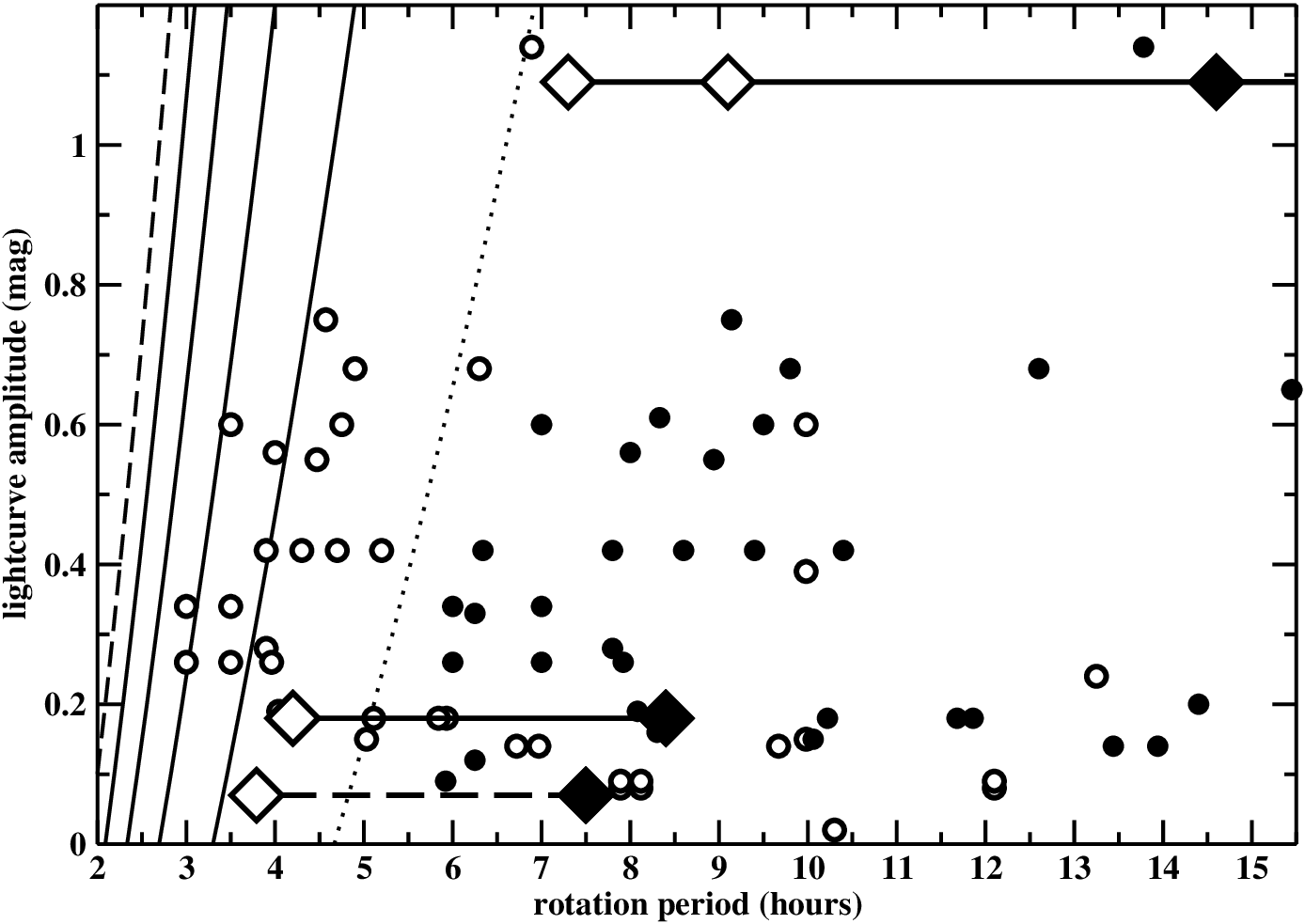}
\caption[]{{\small Amplitudes and
periods for all KBOs and Centaurs
(excluding comets)
with observed lightcurve variations
(data from Table~\ref{otherdata}).
Large diamonds represent
our HST observations and small
circles are data from
other sources.
Open symbols indicate photometric
periods; filled symbols indicate
rotation periods that are twice
the observed photometric period.
Objects known to have double-peaked
lightcurves are plotted only as filled
symbols.
Horizontal lines connect photometric
and twice photometric
solutions for our
HST observations of \newbf,
\newbf, and \fv\ (the horizontal line
for \fv\ is dashed to suggest that the 
lightcurve likely is double-peaked
and that the 3.79~hour solution is
unlikely);
similar
lines could be drawn for most
observations represented here,
as most KBO lightcurves do not allow
distinguishing between these
two solutions,
but we omit these connecting lines for other data
for clarity.
The curves represent solutions
for critical rotation periods
($P_c$) for various densities
(in g~cm$^{-3}$):
0.5~(dotted line),
1, 1.5, 2, 2.5, and 3~(dashed) (from
Equation~\ref{pcrit}).
Regions above and to the left of
a given curve are unstable for a
given density.
A maximum KBO density of
1.0--1.5~g~cm$^{-3}$ (for single-peaked
solutions) or perhaps 0.5~g~cm$^{-3}$
(for double-peaked solutions)
is suggested by the clustering of
points
up to but not beyond the curves
for those densities.
Compare to Figure~8 of
\citet{ph00}.}
}
\label{spin}
\end{figure}

%
\noindent This relation assumes a fluid body,
that is, $\phi = 0$.
Although more rigorous treatments of 
lightcurve data are possible, as shown above,
we will here make this assumption to
allow ready comparisons among bodies
(and to the main belt asteroid
and NEO populations).
Following \citet{ph00}, we plot lightcurve
amplitudes and observed periods for
all presently known KBO and Centaur
data, including our new HST data
for \newbf , \newbg , and~\fv\
(Figure~\ref{spin}).
The rotation periods of most KBOs
and Centaurs could be either
the observed photometric period
(open circles in Figure~\ref{spin})
or twice the photometric period
(closed symbols). For cases in
which the true periods are known
from double-peaked lightcurves,
only this true period is plotted
(closed symbol).

In Figure~\ref{spin} we also show solutions
corresponding to critical periods
for
densities spanning the range of
plausible values for icy-rocky bodies.
Remarkably, there is an apparent
``rotation rate barrier'' in that
there appear to be no KBOs or Centaurs
whose densities must be greater
than 1~or 1.5~g~cm$^{-3}$; this conclusion
is derived from the case in which
rotation periods are identical
to photometric periods.
Similarly, assuming that 
rotation periods are twice the photometric
periods shows that there are no
KBOs or Centaurs whose densities
must be greater than 0.5~g~cm$^{-3}$.
This does not preclude larger
densities, but means that no
KBOs or Centaurs are observed to
have rotations that
{\em require} larger densities.
Furthermore, \citet{ph00}
interpret their results for
NEOs
by saying that
the density that corresponds to the
``rotation rate barrier'' is likely
the maximum bulk density for that
population. While it seems unlikely
that the maximum density for KBOs 
is less than 1~g~cm$^{-3}$, it is nevertheless
remarkable that no KBOs or Centaurs
require densities larger than 
around 0.5~or 1.5~g~cm$^{-3}$.
For comparison, we note that
(47171) 1999~TC$_{36}$
has a density around
0.5~g~cm$^{-3}$ \citep{stans05};
that
the ``rotation rate barrier'' for 
comets is around 0.6~g~cm$^{-3}$
\citep{weis_comets2}; and that
this ``barrier'' for NEOs
is 2--3~g~cm$^{-3}$ for bodies larger
than 200~m \citep{ph00}.

Since KBOs and Centaurs are expected
to be a mixture of ice (density
around~1~g~cm$^{-3}$) and
rock (density perhaps around~3~g~cm$^{-3}$),
we can roughly estimate that porosity
may be important at the level of
tens of percent (see below).
A further implication is that KBOs and
Centaurs in this size
range generally may not have significant
tensile strength, which would allow
stable KBO solutions
to the upper left of the critical
lines shown in Figure~\ref{spin}
(recall that the discussion in Section~\ref{strength}
refers to cohesionless bodies).
This is further confirmation that KBOs
and Centaurs larger than
25~km diameter
are likely to be rubble
piles.
Our general
conclusion from this analysis --- that the bulk densities
of KBOs and Centaurs
likely lie in the range
0.5--1.5~g~cm$^{-3}$ --- is not
surprising
and confirms results that we have
shown above.

Finally, we note that the percentage of small
KBOs with detected lightcurves is
significantly greater than the percentage
of large KBOs with detected lightcurves (Table~\ref{otherdata}).
This is consistent with the arguments
presented above: more small KBOs are likely
to be fragments than large KBOs; fragments are more likely
to be non-spherical than primordial bodies; lightcurves are likely
to be produced by non-spherical bodies;
therefore, a greater percentage of
small KBOs should show significant lightcurve
variations than large KBOs.
We again restate that the data presented in
Table~\ref{otherdata} is certainly biased against
null results and biased toward the detection of small amplitude
lightcurves for big (but not small) KBOs.
Nevertheless, if taken at 
face value,
the data presented in Table~\ref{otherdata}
therefore supports the theoretical models described
above, with the largest bodies remaining
undisrupted since accretion and smaller bodies
representing collisionally-derived fragments.

\subsection{Comparison to other Solar System bodies}
\label{comparison}

The KBOs and Centaurs shown
in Figure~\ref{spin} are generally
hundreds of kilometers, as is
\fv , but
\newbf\ and \newbg\
have diameters a factor of five smaller:
gravity may be important in rounding
bodies larger than a few hundred
kilometers, but does not prohibit
smaller bodies
from maintaining various
extreme shapes (e.g., \citet{dcr02}).
Therefore, the
same physical processes and interpretations
may not be relevant across size regimes
within the Kuiper Belt,
and it is possible that better analogies
of individual objects
are found elsewhere in the 
Solar System, despite
differing collision rates and
ice/rock fractions.

Outer planet satellites may be useful 
analogies to hundred kilometer KBOs;
indeed, some outer planet satellites
may be captured KBOs \citep{johnson05}.
Jupiter's moon Amalthea has
$a/b=1.8$, $a/c$ around~2,
and a derived density
of less than~1.0~g cm$^{-3}$
\citep{anderson05}.
(Compare this result to
the plausible solutions for
\fv\ shown in Figure~\ref{fvfig}.)
The best interpretation for this modest-sized body ---
with mean radius around 80~km, Amalthea is very close
in size to \fv\ ---
is a porosity of tens of percent
even when the satellite is largely
water ice.
The physical state of
this body is not presently
understood,
so we can draw no useful analogy from
it,
other than to say that extremely
low densities in the Solar System (including 0.5~g~cm$^3$ for
(47171) 1999~TC$_{36}$ \citep{stans05}) appear
to be just that: extreme, but not
forbidden.

The
maximum asphericity of \fv\
may be only a few percent (barring pole-on
alignment or a
pathological combination of dark
surface regions along the long axis
and bright surface regions along the
short axis of an elongated body).
The size of \fv\ is similar to
a number of outer Solar System moons.
Uranus' moon
Puck's axis ratio is
close to unity \citep{kark},
but all of these other satellites ---
which are presumably captured and
perhaps fragments of
disrupted bodies ---
are known to be at least
10\% aspherically irregular\footnote{Jupiter: Himalia
has $b/a=0.8$ \citep{porco} and
Amalthea has $b/a=0.58$ \citep{thomas98}.
Saturn:
Phoebe, which has a
retrograde orbit possibly implying capture from
the asteroid belt or Kuiper Belt,
is 10\% to 20\% aspherical
\citep{kruse86,bauer04,porco05};
Epimetheus has $b/a=0.80$
while Janus has $b/a=0.98$ and
$c/a=0.79$ \citep{thomas89}. Neptune:
Despina and Galatea have
$b/a=0.82$
and $b/a=0.9$, respectively,
while
Larissa has $b/a=0.94$ but
$c/a=0.78$
\citep{karknept}.},
though we note that viewing geometries
may play some role (outer planet satellites,
except Uranus', tend to be viewed close
to equatorially, maximizing lightcurve
variations, whereas KBOs are assumed
to have randomized obliquities that
are more likely to hide their true shapes).
Furthermore, \citet{sj} compile a list
of aspherical Solar System
objects larger than 200~km and
suggest that the four KBOs they
observed to have lightcurve variations ---
all larger than 200~km ---
may also be irregular, with
asphericities of tens of percent.
It is thus remarkable that
even modest asphericity of the
116~km \fv\ is unlikely based on our
photometry (barring
a pole-on orientation).
Perhaps impacts have more
thoroughly pulverized \fv\ (and KBOs)
than satellites of
giant planets.
\fv\ would therefore have
small internal friction and would be
more relaxed
and closer to the fluid equilibrium
state.
We note that approximately half of the
KBOs that have been searched for
photometric variability show no
such
signal, typically with sensitivities
around 0.1~magnitudes. This 50\% null
result could be interpreted as suggesting
that many hundred kilometer-sized KBOs
are less than 10\% aspherical. A significant
difference between KBOs and outer Solar
System satellites may be implied.

We can look to the comet population for 
relevant analogies for the smaller KBOs.
\citet{jewitt03} studied
shapes of comet nuclei, which are 
an order of magnitude smaller
than the HST KBOs and two orders of
magnitude smaller than most other
well-studied KBOs. They conclude that 
the primary cause of comet nuclei
asphericity likely is extensive mass
loss. We suspect that such a process
is not significant for classical KBOs,
such as the four we observed with HST,
that never
approach closer to the Sun than
$\sim$35~AU,
but could be important for Centaurs,
which can have
semi-axes as small as $\sim$15~AU.

\citet{weis_comets2} compiled 
rotation periods and projected
axis ratios ($a/b$) for 13~short-period
comets
and carried out an analysis similar
to our Section~\ref{densities}
and Figure~\ref{spin}.
They find an apparent
``rotation rate barrier''
that corresponds to 
an upper limit density around
0.6~g~cm$^{-3}$,
similar to the upper limit
we derive from the double-period
solutions for KBOs (filled circles
in Figure~\ref{spin}). Comets
clearly have non-gravitational
forces (e.g., jets) that can affect
both shape and rotation periods, so
this apparent agreement should not
be overemphasized. Nevertheless,
the idea that short-period
(Jupiter-family) comets derive
from the Kuiper Belt (e.g., \citet{levison97})
may be supported
by this agreement.

%
%

Finally, the asteroid belt includes bodies throughout
the size range of KBOs and may prove useful for understanding
the physical properties of KBOs.
\fv\ has no good close analog 
among main belt asteroids (using
absolute magnitude, lightcurve
amplitude, and period as criteria).
However, 
\newbf\ may have a good
and easily imagined analog in the
main asteroid belt, based on 
lightcurve amplitude
and approximate size:
asteroid (243)~Ida,
which has maximum and minimum
dimensions of 55.3~km and
14.6~km, asymmetry (area-weighted
average of the ratio of
the radii) of~1.48,
and an observed 
lightcurve around 0.8~magnitudes 
\citep{simon96,thomas96}.
Ida has clearly been much affected
by disruptive collisions, as 
suggested by its membership in the
Koronis dynamical family; by
the presence of its
(presumably impact-generated) satellite,
Dactyl;
and by its much-cratered appearance
\citep{green96}.
All evidence suggests that Ida is a
collisional fragment of the (former)
Koronis parent body.
Note that Ida's significant aspect ratio
demonstrates, at least in concept, that 
fragmentary results of collisional events
can have substantially aspherical shapes
and consequently large amplitude lightcurves.
We note that the asteroid belt has
a higher space density of bodies
and larger impact speeds than the
Kuiper Belt.
Perhaps, however, it
is not inappropriate to imagine
an icy
Ida when picturing \newbf .

%

\section{Conclusions}
\label{conclusions}

We have derived best-fit lightcurves for
four KBOs imaged in the HST/ACS KBO
survey \citep{gmb}.
\newbf\ is found to experience
large amplitude 
periodic brightness variations,
whereas \fv\ significantly is found
to undergo very small but non-zero amplitude
periodic brightness variations that are
non-sinusoidal.
Our primary conclusions are the following:

\begin{itemize}
\item[(1)]
Plausibly, based on the 
range of suggested and measured albedos for KBOs,
an albedo range of at
least a factor of~2.5 could exist on \newbf,
although
such unlikely and 
extreme albedo ranges on single
bodies in the outer Solar System
are seen only
in unusual
situations.
However,
albedo and shape could
be correlated, as would be the case
with a large, fresh, bright crater.  Furthermore it may be easier for
a crater or albedo feature to dominate the majority of a hemisphere of a
small body like \newbf , so we cannot
exclude the possibility of a wide range of surface reflectance on
\newbf.

\item[(2)]  \newbf\ 
could have
complicated topography that produces
lightcurves that --- at least during
our observing season --- are substantially
larger than their gross shapes might
otherwise indicate.
Facets on \fv\ could produce a 
small amplitude lightcurve that suggests
a body more spherical than its true
shape.
Additionally, the relatively small
deviations from sphericity required
to produce the observed \fv\ lightcurve
may be readily explained by topography -- facets -- 
in the presence of low surface gravity.

\item[(3)]
The conditions in which small
amplitude lightcurves are produced
(e.g., \fv )
include bodies of any shape
seen nearly pole-on ($\theta \approx 0$)
and
nearly spherical bodies
($a\approx b\approx c$)
seen at any angle.
For Jacobian solutions --- the only non-pole-on
fluid solutions
in which the photometric lightcurve is derived
from the gross aspherical shape of the body --- the bulk density of \fv\
must be
0.67--1.0~g~cm$^{-3}$.
For Maclaurin solutions (rotating spot
model) as well as for pole-on orientations,
the minimum density is
2.7~g~cm$^{-3}$.

The simplest solution arises from allowing
non-zero internal friction:
\fv\ can readily be a rubble pile
with density 1--2~g~cm$^{-3}$
and small (but non-zero)
internal friction.

\end{itemize}

The emerging picture for \fv\ is
that of a body that accreted to approximately
its present size; has been completely
shattered due to extensive collisions;
has little internal friction due to
its rubble pile nature ($\phi$ small
but likely non-zero); and is nearly,
but not completely, relaxed, thus
nearly attaining a rotating
fluid equilibrium
state.
This conclusion is consistent with the 
idea that the timescale for disruptive
collisions among 100~km
KBOs is longer than
the Solar System.
The non-sinusoidal lightcurve could
be produced by facets or surface
topography, or simply as a result
of \fv\ being nearly, but not quite,
in rotational fluid equilibrium.

\newbf\ (as well as \newbg ) is likely a single
coherent fragment, the result of an
unusual, but not wildly improbable,
collision between 100~kilometer bodies.

We combine the new lightcurve data presented
here 
with all other reported KBO photometry
to understand
the physical properties of the KBO
population.
Our general
conclusion from this analysis is
that the bulk densities
of KBOs and Centaurs
likely lie in the range
0.5--1.5~g~cm$^{-3}$.
This is consistent with the results of
the detailed modeling we carried out
for the HST/ACS KBOs
and roughly consistent with the
average bulk density for short-period
comets. This agreement may strengthen
the proposed genetic link between
Kuiper Belt Objects and short-period
comets.
We furthermore show that the percentage
of small KBOs with lightcurve variations
is greater than that for large KBOs, implying
that small KBOs are non-spherical fragments
produced by collisions.

Outer Solar System satellites of 
the size of \fv\ 
almost all have asphericities
greater than 10\%.
Perhaps 50\% of 
similarly-sized KBOs show no
variability at the 10\% level,
suggesting
a significant
difference between the evolutions
of KBOs and outer Solar 
System satellites.

The most helpful and easily
imagined Solar System analog 
for \newbf\
may be
the main
belt asteroid (243)~Ida,
which has size, axis ratios, and shape
that are similar to those we derive
for \newbf.
Ida has clearly been much affected
by disruptive collisions and is a fragment
of a larger parent body,
further suggesting that
\newbf\ could be a collisionally shaped
body.
Perhaps it
is not inappropriate to imagine
an icy
Ida when picturing the
small KBO \newbf .

\acknowledgments We thank Erik Asphaug, Derek
Richardson, and Keith Holsapple
for enlightening conversations
about strength.
Larry Wasserman helped us
mine the Lowell
asteroid database for KBO analogs.
Our computer cluster at Penn
is expertly maintained by
Matt Lehner
and Rahul Dave.
Tony Roman
and Ron Gilliland at STScI helped
with the detailed planning and execution
of the HST ACS KBO survey. 
Al Harris provided useful insight
about lightcurves and NEOs, and we thank
an anonymous referee for a careful
reading and good suggestions.

This work
was supported by STScI grant
GO-9433.06.
Support for program \#GO-9433 was provided
by NASA through a grant from
the Space Telescope Science Institute,
which is operated by the Association of
Universities for Research in 
Astronomy, Inc., under NASA 
contract NAS~5-26555.

\end{document}